\documentclass[twocolumn,floatfix,showpacs,prd,aps,tightenlines]{revtex4}
\usepackage{graphicx}
\usepackage{amsmath}
\usepackage{psfrag}
\usepackage{dcolumn}
\usepackage{bm}

\newcommand{\bea}{\begin{eqnarray}}
\newcommand{\eea}{\end{eqnarray}}
\newcommand{\beq}{\begin{equation}}
\newcommand{\eeq}{\end{equation}}
\newcommand{\KMS}{\rm km\,s^{-1}}
\newcommand{\vz}{v_\|}

\begin{document}

\title{Modeling gravitational recoil from precessing highly-spinning
unequal-mass black-hole binaries}

\author{Carlos O. Lousto}
\affiliation{Center for Computational Relativity and Gravitation,
School of Mathematical Sciences,
Rochester Institute of Technology, 78 Lomb Memorial Drive, Rochester,
 New York 14623}

\author{Yosef Zlochower} 
\affiliation{Center for Computational Relativity and Gravitation,
School of Mathematical Sciences,
Rochester Institute of Technology, 78 Lomb Memorial Drive, Rochester,
 New York 14623}

\date{\today}

\begin{abstract}

We measure the gravitational recoil  for
unequal-mass-black-hole-binary mergers, with the larger BH having spin
$a/m^H=0.8$, and the smaller BH non-spinning. We choose our
configurations such that, initially, the spins lie on the orbital
plane. The spin and orbital plane precess significantly, and we find
that the out-of plane recoil (i.e.\ the recoil perpendicular to the
orbital plane around merger) varies as $\eta^2 / (1+q)$, in agreement
with our previous prediction, based on the post-Newtonian scaling.
\end{abstract}

 \maketitle

\section{Introduction}

The recent observational discovery of a possible recoiling
supermassive black hole at a speed of $2650\ \KMS$ with respect to its
host galaxy~\cite{Komossa:2008qd} represent the first observational
evidence in support of the predictions of General Relativity in the
strong-field, highly-dynamical, and highly-nonlinear regime. This
recoil, if it in fact resulted from a black-hole merger, would confirm
the theoretical prediction of Campanelli et
al.~\cite{Campanelli:2007ew} that black-hole mergers can lead to very
large recoils.
This original prediction,  and the subsequent
calculation~\cite{Campanelli:2007cga} that indicated that such recoils
can be as large as $4000\ \KMS$, has been a key trigger for the recent
theoretical efforts to determine EM signatures of large recoils and
the astronomical searches for these
signatures~\cite{Bogdanovic:2007hp, Loeb:2007wz, Bonning:2007vt,
HolleyBockelmann:2007eh, Komossa:2008qd, Komossa:2008ye}.

Thanks to recent breakthroughs in the full non-linear numerical
evolution of black-hole-binary spacetimes~\cite{Pretorius:2005gq,
Campanelli:2005dd, Baker:2005vv}, it is now possible to accurately
simulate the merger process and examine its effects in this highly
non-linear regime~\cite{Campanelli:2006gf, Baker:2006yw,
Campanelli:2006uy, Campanelli:2006fg, Campanelli:2006fy,
Pretorius:2006tp, Pretorius:2007jn, Baker:2006ha, Bruegmann:2006at,
Buonanno:2006ui, Baker:2006kr, Scheel:2006gg, Baker:2007fb,
Marronetti:2007ya, Pfeiffer:2007yz}.  Black-hole binaries will radiate
between $2\%$ and $8\%$ of their total mass and up to $40\%$ of their
angular momenta, depending on the magnitude and direction of the  spin
components, during the merger~\cite{Campanelli:2006uy,
Campanelli:2006fg, Campanelli:2006fy}.  In addition, the radiation of
net linear momentum by a black-hole binary leads to the recoil of the
final remnant hole~\cite{Campanelli:2004zw,  Herrmann:2006ks,
Baker:2006vn, Sopuerta:2006wj, Gonzalez:2006md, Sopuerta:2006et,
Herrmann:2006cd, Herrmann:2007zz, Herrmann:2007ac, Campanelli:2007ew,
Koppitz:2007ev, Choi:2007eu, Gonzalez:2007hi, Baker:2007gi,
Campanelli:2007cga, Berti:2007fi, Tichy:2007hk, Herrmann:2007ex,
Brugmann:2007zj, Schnittman:2007ij, Krishnan:2007pu,
HolleyBockelmann:2007eh, Pollney:2007ss, Dain:2008ck}, which can have
astrophysically important effects \cite{Redmount:1989, Merritt:2004xa,
Campanelli:2007ew, Gualandris:2007nm, HolleyBockelmann:2007eh,
Kapoor76}.

In~\cite{Campanelli:2007ew} we introduced the following 
heuristic model for the
gravitational recoil of a merging binary.
\begin{equation}\label{eq:empirical}
\vec{V}_{\rm recoil}(q,\vec\alpha_i)=v_m\,\hat{e}_1+
v_\perp(\cos(\xi)\,\hat{e}_1+\sin(\xi)\,\hat{e}_2)+\vz\,\hat{e}_z,
\end{equation}
where
\begin{subequations}
\begin{equation}\label{eq:vm}
v_m=A\frac{\eta^2(1-q)}{(1+q)}\left(1+B\,\eta\right),
\end{equation}
\begin{equation}\label{eq:vperp}
v_\perp=H\frac{\eta^2}{(1+q)}\left(\alpha_2^\|-q\alpha_1^\|\right),
\end{equation}
\begin{equation}\label{eq:vpar}
\vz=K\cos(\Theta-\Theta_0)\frac{\eta^2}{(1+q)}\left|\vec\alpha_2^\perp-q\vec\alpha_1^\perp\right|,
\end{equation}
\end{subequations}
$A = 1.2\times 10^{4}\ \KMS$~\cite{Gonzalez:2006md},
$B = -0.93$~\cite{Gonzalez:2006md},
$H = (6.9\pm0.5)\times 10^{3}\ \KMS$,
$\vec{\alpha}_i=\vec{S}_i/m_i^2$,
$\vec S_i$ and $m_i$ are the spin and mass of
hole $i$, $q=m_1/m_2$ is the mass ratio of the smaller to larger mass hole,
$\eta = q/(1+q)^2$ is the symmetric mass ratio,
the index $\perp$ and $\|$
refer to perpendicular and parallel to the orbital angular momentum
respectively at the effective moment of the maximum generation of
the recoil (around merger time),
$\hat{e}_1,\hat{e}_2$ are orthogonal unit vectors in the
orbital plane, and $\xi$ measures the angle between the ``unequal mass''
and ``spin'' contributions to the recoil velocity in the orbital plane.
The angle $\Theta$ was defined as the angle between the in-plane
component of $\vec \Delta\equiv (m_1+m_2)({\vec S_2}/m_2 -{\vec S_1}/m_1)$ 
and the infall direction at merger.
The form of Eq.~(\ref{eq:vm}) was proposed 
in~\cite{1983MNRAS.203.1049F,Gonzalez:2006md},
while the form of Eqs.~(\ref{eq:vperp})~and~(\ref{eq:vpar}) was 
proposed in~\cite{Campanelli:2007ew}
based on the post-Newtonian expressions in~\cite{Kidder:1995zr}.
 In Ref~\cite{Campanelli:2007cga} we 
determined that
$K=(6.0\pm0.1)\times 10^4\ \KMS$, and made the first prediction that
the maximum possible recoil is $\sim4000\ \KMS$ for equal-mass binaries
with anti-parallel spins in the orbital plane (in
Ref.~\cite{Dain:2008ck}, we performed simulations with a measured
recoil of $3250\ \KMS$).
Although $\xi$ may in general depend strongly on the configuration,
the results of~\cite{Choi:2007eu} and post-Newtonian~\cite{Kidder:1995zr}
 calculations show that $\xi$ is
$90^\circ$ for headon collisions, and the results presented in
Ref.~\cite{Lousto:2007db}
indicate that $\xi \sim 145^\circ$ for a wide range of quasi-circular
configurations.
A simplified version of Eq.~(\ref{eq:empirical}) that
models the magnitude of $V_{\rm recoil}$ was independently proposed
in~\cite{Baker:2007gi}, and a simplified form of Eq.~(\ref{eq:vperp})
for the equal-mass aligned spin case was proposed in~\cite{Koppitz:2007ev}.
A more general formula, using only symmetry arguments and fits to
numerical data, was recently proposed in~\cite{Boyle:2007sz,
Boyle:2007ru}.

Our heuristic formula~(\ref{eq:empirical}) describing the recoil
velocity of a black-hole binary remnant as a function of the
parameters of the individual holes has been theoretically verified in
several ways. In~\cite{Campanelli:2007cga} the $\cos{\Theta}$
dependence was established and was confirmed in~\cite{Brugmann:2007zj}
for binaries with different initial separations. In
Ref.~\cite{Herrmann:2007ex} the decomposition into spin components
perpendicular and parallel to the orbital plane was verified, and
in~\cite{Pollney:2007ss} it was found that the quadratic-in-spin
corrections to the in-plane recoil velocity are less than $20\ \KMS$.
Recently, Baker et al.~\cite{Baker:2008md} measured the recoil for
unequal-mass, spinning binaries, with spins lying in the initial orbital
plane, and concluded that the leading order dependence of the
out-of-plane kick was ${\cal O}(\eta^3)$, rather than the ${\cal
O}(\eta^2)$ that we predicted. In this paper we examine this
dependence in detail.

As pointed out in~\cite{Baker:2008md} the consequences of an ${\cal
O}(\eta^3)$ dependence of the recoil, rather than an ${\cal
O}(\eta^2)$ dependence, are significant for both the retention of
intermediate mass black holes (IMBH) in globular clusters and
supermassive black holes in galaxies. 
It is thus important that we understand how the recoil depends on mass
ratio.

In their paper, Baker et al.~\cite{Baker:2008md} analyzed
configurations that required fitting two angle parameters (for a given
value of $q$) before the maximum recoil could be obtained. Their
configuration also produced very small recoil velocities for smaller
values of $q$. In order to help clearly display the dependence, we
choose configurations that only require fitting one angle parameter in
Eq.~(\ref{eq:vpar}), and have a substantial recoil even for our
smallest mass ratios.  In addition, the Baker et
al.~\cite{Baker:2008md} runs were constructed such that the spin of
the larger black hole is proportional to the mass ratio squared times
the spin of the smaller BH, i.e.\ $\alpha_2 = q^2 \alpha_1$, where
$q=m_1/m_2 < 1$, and $\alpha_1 < 1$ is the spin of the smaller black
hole. In the small mass ratio limit, this leads to a small, very
highly rotating hole (for any spin-induced recoil to be observable)
orbiting a large, essentially nonrotating, BH. This configuration does
not match the expected astrophysical scenario since large BHs are
expected to have high spins. Our choice of configuration more closely
matches the astrophysical scenario when $q\to0$ because in that limit
the recoil becomes independent of the spin of the smaller BH, as is
apparent in the factor of $\alpha_2 - q \alpha_1$ in
Eq.~(\ref{eq:vpar}).

The paper is organized as follows, in Sec.~\ref{sec:techniques}
we
review the numerical techniques used for the evolution of
the black-hole binaries and the analysis of the physical
quantities extracted at their horizons, in Sec.~\ref{sec:res}
we present results and analysis, and in Sec.~\ref{sec:discussion}
we present our conclusions.

\section{Techniques}
\label{sec:techniques}

To compute initial data, we use the puncture approach~\cite{Brandt97b} 
along with the {\sc
TwoPunctures}~\cite{Ansorg:2004ds} thorn.  In
this approach the 3-metric on the initial slice has the form
$\gamma_{a b} = (\psi_{BL} + u)^4 \delta_{a b}$, where $\psi_{BL}$ is
the Brill-Lindquist conformal factor, $\delta_{ab}$ is the Euclidean
metric, and $u$ is (at least) $C^2$ on the punctures.  The
Brill-Lindquist conformal factor is given by
$
\psi_{BL} = 1 + \sum_{i=1}^n m_{i}^p / (2 |\vec r - \vec r_i|),
$
where $n$ is the total number of `punctures', $m_{i}^p$ is the mass
parameter of puncture $i$ ($m_{i}^p$ is {\em not} the horizon mass
associated with puncture $i$), and $\vec r_i$ is the coordinate location of
puncture $i$.  We evolve these black-hole-binary data-sets using the
{\sc LazEv}~\cite{Zlochower:2005bj} implementation of the moving
puncture approach~\cite{Campanelli:2005dd,Baker:2005vv}.  In our version of the
moving puncture approach we replace the
BSSN~\cite{Nakamura87,Shibata95, Baumgarte99} conformal exponent
$\phi$, which has logarithmic singularities at the punctures, with the
initially $C^4$ field $\chi = \exp(-4\phi)$.  This new variable, along
with the other BSSN variables, will remain finite provided that one
uses a suitable choice for the gauge. An alternative approach uses
standard finite differencing of $\phi$~\cite{Baker:2005vv}. 
Recently Marronetti et al.~\cite{Marronetti:2007wz} proposed the use
of $W=\sqrt{\chi}$ as an evolution variable.
For the runs presented here we use centered, eighth-order finite differencing
in space~\cite{Lousto:2007rj} and an RK4 time integrator (note that we
do not upwind the advection terms).

We use the {\sc CACTUS} framework~\cite{cactus_web} with the 
{\sc CARPET}~\cite{Schnetter-etal-03b} mesh refinement
driver to provide a `moving boxes' style mesh refinement. In this
approach  refined grids of fixed size are arranged about the
coordinate centers of both holes.  The {\sc CARPET} code then moves these
fine grids about the computational domain by following the
trajectories of the two black holes.

We obtain accurate, convergent waveforms and horizon parameters by
evolving this system in conjunction with a modified 1+log lapse and a
modified Gamma-driver shift
condition~\cite{Alcubierre02a,Campanelli:2005dd}, and an initial lapse
$\alpha(t=0) = 2/(1+\psi_{BL}^{4})$.
The lapse and shift are evolved with
\begin{subequations}
\label{eq:gauge}
\begin{eqnarray}
(\partial_t - \beta^i \partial_i) \alpha &=& - 2 \alpha K\\
 \partial_t \beta^a &=& B^a \\
 \partial_t B^a &=& 3/4 \partial_t \tilde \Gamma^a - \sigma B^a.
 \label{eq:Bdot}
\end{eqnarray}
\end{subequations}
Note that we denote the Gamma-driver parameter by $\sigma$ rather than
the more typical $\eta$ to avoid confusion with the symmetric mass
ratio parameter.  These gauge conditions require careful treatment of
$\chi$, the inverse of the three-metric conformal factor, near the
puncture in order for the system to remain
stable~\cite{Campanelli:2005dd,Campanelli:2006gf,Bruegmann:2006at}.
In our tests, $W$ showed better behavior at very early times ($t <
10M$) (i.e.\ did not require any special treatment near the
punctures), but led to evolutions with lower effective resolution when
compared to $\chi$. Interestingly, a mixed evolution system that
evolved $W$ for $t < 10 M$ and $\chi$ for $t > 10M$ showed
inaccuracies similar to the pure $W$ system. At higher resolution $W$
and $\chi$ agreed with good accuracy.  In Ref.~\cite{Gundlach:2006tw}
it was shown that this choice of gauge leads to a strongly hyperbolic
evolution system provided that the shift does not become too large.

We use {\sc AHFinderdirect}~\cite{Thornburg2003:AH-finding} to locate
apparent horizons.
We measure the magnitude of the horizon spin using the Isolated
Horizon algorithm detailed in~\cite{Dreyer02a}. This algorithm is
based on finding an approximate rotational Killing vector (i.e.\ an
approximate rotational symmetry) on the horizon, and given this
approximate Killing vector $\varphi^a$, the spin magnitude is
\begin{equation}\label{isolatedspin}
S_{[\varphi]} = \frac{1}{8\pi}\oint_{AH}(\varphi^aR^bK_{ab})d^2V
\end{equation}
where $K_{ab}$ is the extrinsic curvature of the 3D-slice, $d^2V$ is the
natural volume element intrinsic to the horizon, and $R^a$ is the
outward pointing unit vector normal to the horizon on the 3D-slice.
We measure the
direction of the spin by finding the coordinate line joining the poles
of this Killing vector field using the technique introduced
in~\cite{Campanelli:2006fy}.  Our algorithm for finding the poles of
the Killing vector field has an accuracy of $\sim 2^\circ$
(see~\cite{Campanelli:2006fy} for details). The mass of the horizon
is given by the Christodoulou formula
\begin{equation}
{m^H} = \sqrt{m_{\rm irr}^2 +
 S^2/(4 m_{\rm irr}^2)},
\end{equation}
where $m_{\rm irr}$ is the irreducible mass.

We also use an alternative quasi-local measurement of the spin and
linear momentum of the individual black holes in the binary that is
based on the
coordinate rotation and translation vectors~\cite{Krishnan:2007pu}.
In this approach the spin components of the horizon are given by
\begin{equation}
  S_{[i]} = \frac{1}{8\pi}\oint_{AH} \phi^a_{[i]} R^b K_{ab} d^2V,
  \label{eq:coordspin}
\end{equation}
where 
 $\phi^i_{[\ell]} = \delta_{\ell j} \delta_{m k} r^m \epsilon^{i j k}$,
and $r^m = x^m - x_0^m$ is the coordinate displacement from the centroid
of the hole,
while the linear momentum is given by
\begin{equation}
  P_{[i]} = \frac{1}{8\pi}\oint_{AH} \xi^a_{[i]} R^b (K_{ab} - K \gamma_{ab}) d^2V,
  \label{eq:coordmom}
\end{equation}
where 
 $\xi^i_{[\ell]} = \delta^i_\ell$.

We measure radiated energy, linear momentum, and angular momentum, in
terms of $\psi_4$, using the formulae provided in
Refs.~\cite{Campanelli99,Lousto:2007mh}. However, rather than using
the full $\psi_4$ we decompose it into $\ell$ and $m$ modes and solve
for the radiated linear momentum, dropping terms with $\ell \geq 5$
(only the $\ell=2$ and $\ell=3$ modes make significant contributions
to the recoil). The
formulae in Refs.~\cite{Campanelli99,Lousto:2007mh} are valid at $r=\infty$. We obtain highly
accurate values for these quantities by solving for them on spheres of
finite radius (typically $r/M=50, 60, \cdots, 100$), fitting the results to
a polynomial dependence in $l=1/r$, and extrapolating to $l=0$. We
perform fits based on a linear and quadratic dependence on $l$, and
take the final values to be the quadratic extrapolation
with the differences being the extrapolation error.

\subsection{Initial Data}
\label{sec:ID}

We evolve quasi-circular configurations with a more massive spinning
black hole, with specific spin $a/m^H=0.8$ pointing in the initial
orbital plane, and a non-spinning smaller BH. The orbital parameters
where chosen using 3PN parameters for quasi-circular orbits with
orbital period $M \omega = 0.05$, which
provides the puncture locations and momenta.  We normalize the
puncture mass parameters so that the total ADM mass is $1M$ and the
mass ratio is the specified one. We then modify the configurations by
rotating the initial spin direction (which has a small effect on the
total ADM mass). We denote these configurations by QXXXTHYYY, where XXX
denotes the mass ratio (XXX=100 for $q=1$, XXX=66 for $q=2/3$, XXX=50
for $q=1/2$, XXX=40 for $q=1/2.5$, XXX=33 for $q=1/3$,
XXX=25 for $q=1/4$, XXX=17 for $q=1/6$, and XXX=13 for $q=1/8$)
 and YYY gives the angle (in degrees) between the initial
spin direction and the $y$-axis. The initial data parameters are
summarized in Table~\ref{tab:ID}. These configuration have several
advantages when modeling the out-of-plane kick as a function of mass
ratio. First, one need only fit to two parameters (one angle and one
amplitude) to determine the
maximum recoil for a given mass ratio; improving the statistical
reliability of the estimated maximum recoil. Second, the maximum recoil
velocity is quite large, even for $q=1/6$, ensuring that errors
in measuring the maximum recoil velocity are not a significant fraction of the
recoil itself. In addition the functional form used in our non-linear
fits, $A \cos (\vartheta - B)$,  yields a more robust measurement of
$A$ when compared to fits of $C \cos(\vartheta_1 -D) +
E \cos(\vartheta_2-F)$ (as used in Baker et al.) for small
sample sizes (i.e.\ the functional form used here is more amenable to
an accurate fit when the sample size is small). (See
Fig.~\ref{fig:theta_fits})

\begin{widetext}

\begin{table}
\caption {Initial data parameters for the quasi-circular
configurations with a non-spinning smaller mass black hole (labeled 1),
and a larger spinning black hole (labeled 2). The punctures are located
at $\vec r_1 = (x_1,0,0)$ and $\vec r_2 = (x_2,0,0)$, with momenta
$P=\pm (0, P,0)$, spins $\vec S_2 = (S_x, S_y, 0)$, mass parameters
$m^p$, horizon (Christodoulou) masses $m^H$, and total ADM mass
$M_{\rm ADM}$. The configuration are denoted by QXXXTHYYY where XXX
gives the mass ratio (0.17, 0.25, 0.40, 0.33, 0.50, 0.66, 1.00) and YYY
gives the angle in degrees
between the initial spin direction and the $y$-axis. In all cases the
initial orbital period is $M \omega = 0.05$. }
\label{tab:ID}
\begin{ruledtabular}
\begin{tabular}{lccccccccccc}
Config   & $x_1/M$ & $x_2/M$  & $P/M$    & $m^p_1$ & $m^p_2$ & $S_x/M^2$ & $S_y/M^2$ & $m^H_1$ & $m^H_2$ & $M_{\rm ADM}/M$ & $a/m^H$\\
\hline
Q13TH000 & 5.70322 & -0.69841 & 0.054063 & 0.10174 & 0.55179 & 0.00000 & 0.63913 & 0.11114 & 0.88910 & 1.00000 & 0.8086 \\
Q13TH090 & 5.70322 & -0.69841 & 0.054063 & 0.10174 & 0.55179 &-0.63913 & 0.00000 & ****** & 0.88909 & 1.00005 & 0.8086 \\
Q13TH130 & 5.70322 & -0.69841 & 0.054063 & 0.10174 & 0.55179 &-0.48961 &-0.41083 & ****** & 0.88909 & 1.00003 & 0.8086 \\
Q13TH210 & 5.70322 & -0.69841 & 0.054063 & 0.10174 & 0.55179 & 0.31957 &-0.55351 & ****** & 0.88909 & 1.00001 & 0.8086 \\
Q13TH315 & 5.70322 & -0.69841 & 0.054063 & 0.10174 & 0.55179 & 0.45194 & 0.45194 & ****** & 0.88909 & 1.00003 & 0.8086 \\
\\
Q17TH000 & 5.52270 & -0.90366 & 0.066761 & 0.13153 & 0.53157 & 0.00000 & 0.59594 & 0.14310 & 0.85859 & 1.00000 & 0.8084\\
Q17TH090 & 5.52270 & -0.90366 & 0.066761 & 0.13153 & 0.53157 &-0.59594 & 0.00000 & ****** & 0.85858 & 1.00004 & 0.8084\\
Q17TH130 & 5.52270 & -0.90366 & 0.066761 & 0.13153 & 0.53157 &-0.45651 &-0.38306 & ****** & 0.85858 & 1.00002 & 0.8084 \\
Q17TH210 & 5.52270 & -0.90366 & 0.066761 & 0.13153 & 0.53157 & 0.29797 &-0.51610 & ****** & 0.85858 & 1.00001 & 0.8084\\
Q17TH315 & 5.52270 & -0.90366 & 0.066761 & 0.13153 & 0.53157 & 0.42139 & 0.42139 & ****** & 0.85858 & 1.00002 & 0.8084\\
\\
Q25TH000 & 5.18788 & -1.27783 & 0.086696 & 0.18588 & 0.49511 & 0.00000 & 0.52144 & 0.20081 & 0.80326 & 1.00000 & 0.8082 \\
Q25TH090 & 5.18788 & -1.27783 & 0.086696 & 0.18588 & 0.49511 &-0.52144 & 0.00000 & 0.20094 & 0.80324 & 1.00000 & 0.8082 \\
Q25TH130 & 5.18788 & -1.27783 & 0.086696 & 0.18588 & 0.49511 &-0.39945 &-0.33518 & 0.20080 & 0.80325 & 1.00000 & 0.8082 \\
Q25TH210 & 5.18788 & -1.27783 & 0.086696 & 0.18588 & 0.49511 & 0.26072 &-0.45158 & 0.20081 & 0.80325 & 1.00000 & 0.8082 \\
Q25TH315 & 5.18788 & -1.27783 & 0.086696 & 0.18588 & 0.49511 & 0.36871 & 0.36871 & 0.20081 & 0.80325 & 1.00000 & 0.8032 \\
\\
Q33TH000 & 4.88556 & -1.60904 & 0.101152 & 0.23415 & 0.46316 & 0.00000 & 0.45982 & 0.25147 & 0.75443 & 1.00000 & 0.8079 \\
Q33TH090 & 4.88556 & -1.60904 & 0.101152 & 0.23415 & 0.46316 &-0.45982 & 0.00000 & 0.25144 & 0.75437 & 0.99994 & 0.8080 \\
Q33TH130 & 4.88556 & -1.60904 & 0.101152 & 0.23415 & 0.46316 &-0.35224 &-0.29557 & 0.25145 & 0.75440 & 0.99996 & 0.8080 \\
Q33TH210 & 4.88556 & -1.60904 & 0.101152 & 0.23415 & 0.46316 & 0.22991 &-0.39822 & 0.25162 & 0.75443 & 0.99998 & 0.8079 \\
Q33TH315 & 4.88556 & -1.60904 & 0.101152 & 0.23415 & 0.46316 & 0.32514 & 0.32514 & 0.25145 & 0.75441 & 0.99997 & 0.8080 \\
\\
Q40TH000 & 4.66490 & -1.84716 & 0.109810 & 0.26900 & 0.44033 & 0.00000 & 0.41792 & 0.28772 & 0.71930 & 1.00000 & 0.8077 \\
Q40TH090 & 4.66490 & -1.84716 & 0.109810 & 0.26900 & 0.44033 &-0.41792 & 0.00000 & 0.28771 & 0.71927 & 0.99990 & 0.8079 \\
Q40TH130 & 4.66490 & -1.84716 & 0.109810 & 0.26900 & 0.44033 &-0.32014 &-0.26863 & 0.28771 & 0.71927 & 0.99994 & 0.8078 \\
Q40TH210 & 4.66490 & -1.84716 & 0.109810 & 0.26900 & 0.44033 & 0.20896 &-0.36193 & 0.28771 & 0.71929 & 0.99997 & 0.8077 \\
Q40TH315 & 4.66490 & -1.84716 & 0.109810 & 0.26900 & 0.44033 & 0.29551 & 0.29551 & 0.28771 & 0.71929 & 0.99995 & 0.8078 \\
\\
Q50TH000 & 4.36532 & -2.16588 & 0.119233 & 0.31591 & 0.40994 & 0.00000 & 0.36487 & 0.33611 & 0.67223 & 1.00001 & 0.8074 \\
Q50TH090 & 4.36532 & -2.16588 & 0.119233 & 0.31591 & 0.40994 &-0.36487 & 0.00000 & 0.33609 & 0.67217 & 0.99984 & 0.8076 \\
Q50TH130 & 4.36532 & -2.16588 & 0.119233 & 0.31591 & 0.40994 &-0.27950 &-0.23453 & 0.33611 & 0.67220 & 0.99991 & 0.8076 \\
Q50TH210 & 4.36532 & -2.16588 & 0.119233 & 0.31591 & 0.40994 & 0.18243 &-0.31598 & 0.33611 & 0.67222 & 0.99997 & 0.8074 \\
Q50TH315 & 4.36532 & -2.16588 & 0.119233 & 0.31591 & 0.40994 & 0.25800 & 0.25800 & 0.33613 & 0.67221 & 0.99993 & 0.8075 \\
\\
Q66TH000 & 3.93712 & -2.61286 & 0.128421 & 0.38234 & 0.36743 & 0.00000 & 0.29620 & 0.40390 & 0.60587 & 1.00000 & 0.8069 \\
Q66TH090 & 3.93712 & -2.61286 & 0.128421 & 0.38234 & 0.36743 &-0.29620 & 0.00000 & 0.40389 & 0.60579 & 0.99975 & 0.8072 \\
Q66TH130 & 3.93712 & -2.61286 & 0.128421 & 0.38234 & 0.36743 &-0.22690 &-0.19039 & 0.40388 & 0.60582 & 0.99985 & 0.8071 \\
Q66TH210 & 3.93712 & -2.61286 & 0.128421 & 0.38234 & 0.36743 & 0.14810 &-0.25651 & 0.40389 & 0.60585 & 0.99994 & 0.8070 \\
Q66TH315 & 3.93712 & -2.61286 & 0.128421 & 0.38234 & 0.36743 & 0.20944 & 0.20944 & 0.40389 & 0.60583 & 0.99988 & 0.8071 \\
\\
Q100TH000& 3.28027 & -3.28027 & 0.133568 & 0.48338 & 0.30398 & 0.00000 & 0.20595 & 0.50543 & 0.50547 & 1.00001 & 0.8061 \\
Q100TH090& 3.28027 & -3.28027 & 0.133568 & 0.48338 & 0.30398 &-0.20595 & 0.00000 & 0.50541 & 0.50537 & 0.99965 & 0.8065 \\
Q100TH130& 3.28027 & -3.28027 & 0.133568 & 0.48338 & 0.30398 &-0.15777 &-0.13238 & 0.50542 & 0.50541 & 0.99980 & 0.8063 \\
Q100TH210& 3.28027 & -3.28027 & 0.133568 & 0.48338 & 0.30398 & 0.10297 &-0.17836 & 0.50543 & 0.50544 & 0.99992 & 0.8062 \\
Q100TH315& 3.28027 & -3.28027 & 0.133568 & 0.48338 & 0.30398 & 0.14563 & 0.14563 & 0.50542 & 0.50542 & 0.99983 & 0.8063 \\

\end{tabular}
\end{ruledtabular}
\end{table}
\end{widetext}

\subsection{Determining the Orbital Plane}
\label{sec:orbital_plane}
\begin{figure}
\includegraphics[width=3.5in]{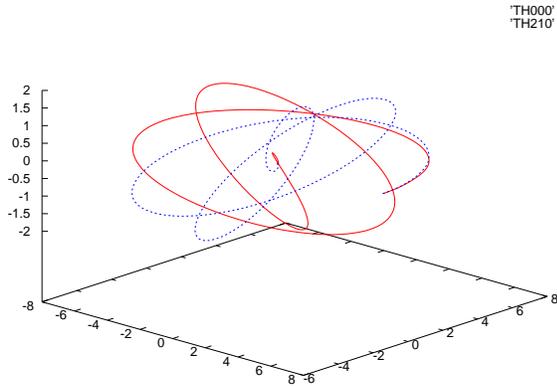}
\caption{The trajectory $\vec r_1 - \vec r_2$ for the Q25TH000 and
Q25TH210 configurations. Note the significant precession and the lack
of alignment between the late-time orbital planes.}
\label{fig:Q25_3d_orbit}
\end{figure}

These configuration show significant orbital precession, as
demonstrated in Fig.~\ref{fig:Q25_3d_orbit}, that presents a
significant challenge when modeling the recoil. Our empirical
formula~(\ref{eq:empirical}) decomposes the recoil in terms of
velocities parallel and perpendicular to the angular momentum. Thus we
need an accurate determination of the orbital plane near merger (where
most of the recoil is
generated~\cite{Brugmann:2007zj,Lousto:2007db}). We determine an
approximate plane by choosing points along the plunge
trajectory near where $|\dot r|$ and $|\ddot r|$  have extrema. We then find a
rotation, such that the late-time approximate orbital plane is rotated onto the
$xy$-plane.
 In order to model the out-of-plane recoil, we need to
measure it as a function of the orientation of
the spin vector during merger. We do this by fixing the remaining
freedom in the transformation such that the transformed trajectories 
(for a given sequence of configurations) coincide (approximately)
near merger. We then measure the spin direction at some fixed fiducial
point along the merger trajectory (in practice, at the point $r_{0}$
discussed below) and fit the out-of-plane kick to
the form $V=A \cos(\vartheta - B)$, where $\vartheta$ is the angle between
the in-plane component of the spin at this fiducial point for a given
configuration with the in-plane spin for the corresponding QXXXTH000 configuration.

We determine the orbital plane during merger in the following way.  We
plot $r=|\vec r_1 - \vec r_2|$ as a function of time and determine
the location of the maximum of $|\dot r|$.  We then choose two points on either
side of the maximum with similar values of $|\dot r|$ and a third
point close to the maximum.  We denote these three points of the
trajectory with $\vec r_+$, $\vec r_0$, and $\vec r_-$.  We then
choose a rotation such that the vector $\vec r_a = \vec r_+ - \vec
r_0$ lies on the new $y$-axis and that the normal to the plane determined
by $\vec r_a$ and $\vec r_b = \vec r_- - \vec r_0$ lies along the new
$z$-axis. With these choices we uniquely determine a rotational
transformation from the coordinates used by the code to coordinates
where the plunging orbital plane coincides with the $xy$ plane. This
transformation also rotates the trajectories in the appropriate way
such that trajectories with similar late-time dynamics will overlap
during merger (this is a generalization of the procedure given in
Ref.~\cite{Dain:2008ck,Lousto:2007db}).

\begin{figure}
\includegraphics[width=3.5in]{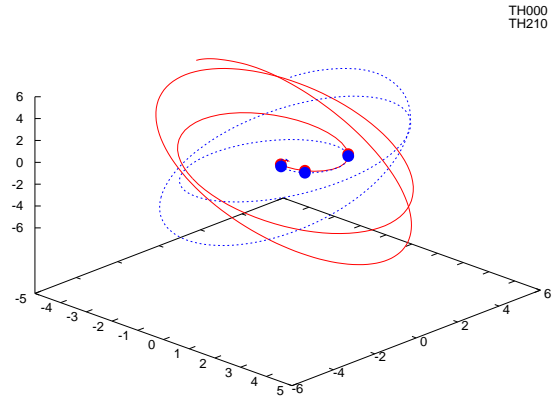}
\caption{The trajectory $\vec r_1 - \vec r_2$ for the Q25TH000 and
Q25TH210 configurations after rotating the system. Note the good
agreement in the late time trajectories and that these trajectories
now lie on the $xy$ plane. The solid points are the locations of
$\vec r_{+}$, $\vec r_{0}$, and $\vec r_{-}$. Note that these points
agree for both curves.}
\label{fig:Q25_3d_orbit_transformed}
\end{figure}

In Fig.~\ref{fig:Q25_3d_orbit} we show the orbital trajectory
difference $\vec r_1 - \vec r_2$ for the Q25TH000 and Q25TH210
configurations. Note the significant precession and that the orbital
planes near merger do not coincide.
Fig.~\ref{fig:Q25_3d_orbit_transformed} shows that, after rotating the
plane, as described above, the merger trajectories coincide. 

\begin{figure}
\includegraphics[width=3.0in]{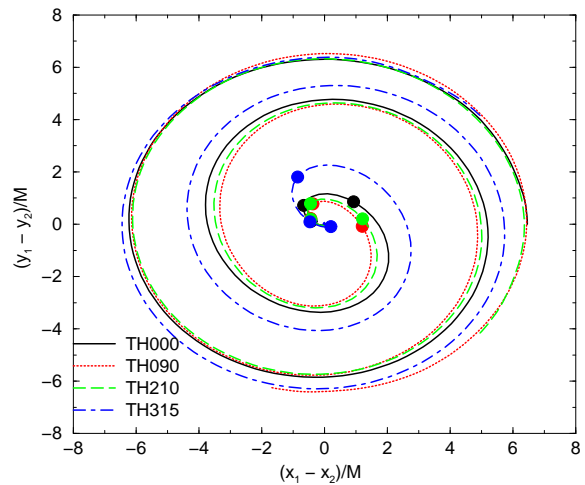}
\caption{The $xy$ projection for the untransformed trajectory $\vec
r_1 - \vec r_2$ for the Q25THYYY configurations. The trajectories have
been rotated (in the $xy$ plane). Note that there are no rotations in
the $xy$ plane that will make the late-time trajectories overlap. The
filled circles are the
locations of $\vec r_{+}$, $\vec r_{0}$, and $\vec r_{-}$.}
\label{fig:Q25_2d_orbit}
\end{figure}

\begin{figure}
\includegraphics[width=3.0in]{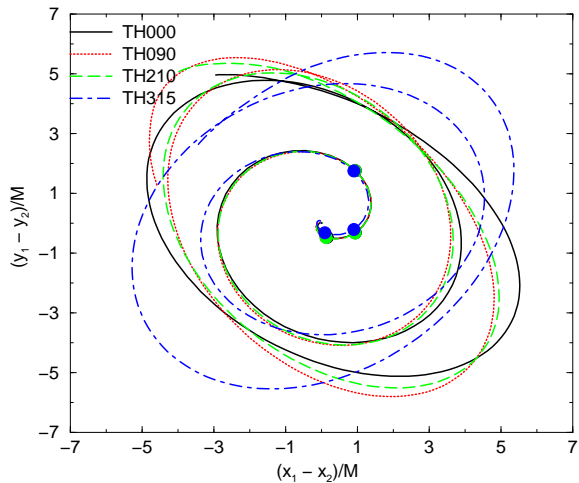}
\caption{The  $xy$ projection for the transformed trajectory $\vec r_1
- \vec r_2$ for the Q25THYYY configurations. No additional rotations
  have been applied.  Note the good
agreement in the late time trajectories. The filled circles are the
locations of $\vec r_{+}$, $\vec r_{0}$, and $\vec r_{-}$.
}
\label{fig:Q25_2d_orbit_transformed}
\end{figure}

In Figs.~\ref{fig:Q25_2d_orbit}~and~\ref{fig:Q25_2d_orbit_transformed}
we show the $xy$ projections of the trajectories for the Q25THYYY
configurations before and after rotating the merger orbital plane.
Note that prior to this transformation, there is no rotation in the
$xy$ plane that will make the late-time trajectories overlap and the
reasonable overlap of the late-time trajectories for the transformed
case.
\section{Results}
\label{sec:res}

We used the following grid configurations for the runs. For the
$q=1/2$, $q=1/2.5$, and $q=1/3$ runs we placed the outer boundaries at
$1664M$ with a coarsest resolution of $h=25.6M$, we used 12 levels of
refinement around the smaller BH, with finest resolution of $h=M/80$,
and 11 levels of refinement around the larger BH. For the $q=1/4$ runs
we added an additional level of refinement around the smaller BH, for
$q=1/6$ and $q=1/8$ we added two additional levels (for a total of
14), and for the $q=2/3$ runs we added an additional level of
refinement about the larger BH. For the $q=1$ runs we used 11 levels
of refinement around the non-spinning hole and 12 levels of refinement
around the spinning hole. In the $q=1/4$ case, we found that using
eighth-order accuracy was critical, as incorrect dynamics (outspiral
rather than inspiral when evolving with $\chi$, and plunges rather
than inspirals when evolving with $W$) resulted when using fourth
order methods. The dynamics obtained from  the $W$ and $\chi$ systems
agreed when using eighth-order methods. In all cases we used
$\sigma=3$ in the Gamma-driver shift condition.  In addition we ran
all $q=1/4$ configuration with a higher resolution (by a factor of
1.2) for all levels (See Sec.~\ref{sec:finite_diff}).

In Table~\ref{table:raw_kick} we give the radiated energy and angular
momenta for all configurations, as well as the untransformed recoil
velocities.
\begin{widetext}

\begin{table}
\caption{The radiated energy and angular momentum, as well as the recoil velocities for each configuration.
Note that some of the error estimates, which are based on the differences between a linear and quadratic
extrapolation in $l=1/r$, are very small. This indicates that the
differences between the extrapolation can underestimate
the true error. All quantities are given in the coordinate system used
by the code (i.e.\ the untransformed system). The Q25THXXX quantities
given on the bottom of the table are for the higher resolution runs.
See Sec.~\ref{sec:finite_diff} for an explanation of the differences
in the radiated quantities.}
\label{table:raw_kick}
\begin{ruledtabular}
\begin{tabular}{lccccccc}

Config   & $100 E/M$ & $100 J_x/M^2$ & $100 J_y/M^2$ & $100 J_z/M^2$ & $V_x$ & $V_y$ & $V_z$\\
\hline
Q13TH000 & $0.623\pm0.018$& $2.383\pm0.031$& $1.987\pm0.050$&$2.002\pm0.035$& $ 66.5\pm  2.3$ & $ 26.9\pm  5.0$ & $101.0\pm  6.8$\\
Q13TH090 & $0.6444\pm0.005$& $-2.182\pm0.008$& $2.533\pm0.017$& $2.142\pm0.184$& $-14.2\pm  6.5$ & $ 70.7\pm  3.3$ & $ 92.7\pm  6.3$\\
Q13TH130 & $0.671\pm0.013$& $-3.438\pm0.034$& $0.509\pm0.008$& $2.112\pm0.012$& $-393.0\pm 11.9$ & $-167.4\pm 15.3$ & $ 43.2\pm  0.1$\\
Q13TH210 &  $0.637\pm0.007$& $-1.072\pm0.031$&$-2.980\pm0.038$& $2.199\pm0.140$& $ 38.1\pm  3.9$ & $ 86.3\pm  2.8$ & $-97.1\pm  7.5$\\
Q13TH315 &  $0.689\pm0.011$& $3.484\pm0.041$&$-0.161\pm0.009$& $2.162\pm0.049$& $-382.0\pm 12.3$ & $-216.2\pm 17.3$ & $-27.2\pm  0.2$\\
\\ \\
Q17TH000 & $1.032\pm0.0343$ & $3.644\pm0.027$ & $3.990\pm0.094$ & $3.70\pm0.094$ & $-99.5\pm 10.8$ & $-624.5\pm 27.6$ & $-137.9\pm  6.8$\\
Q17TH090 & $0.980\pm0.005$ &  $-4.133\pm0.014$ &  $3.903\pm0.078$ & $3.66\pm0.246$ & $323.1\pm  5.2$ & $-105.5\pm  1.8$ & $-146.1\pm 10.0$\\
Q17TH130 & $1.049\pm0.027$ &  $-5.592\pm0.059$ &  $0.004\pm0.014$ &  $3.484\pm0.045$ & $371.8\pm 10.4$ & $343.2\pm 16.4$ & $-42.3\pm  0.3$\\
Q17TH210 & $1.027\pm0.032$ &  $-1.162\pm0.011$ &  $-5.351\pm0.100$ &  $3.70\pm0.077$ & $228.6\pm 21.5$ & $-579.2\pm 14.8$ & $140.8\pm  5.4$\\
Q17TH315 & $1.032\pm0.012$ &  $5.401\pm0.044$ &  $0.352\pm0.035$ &  $3.687\pm0.222$ & $175.0\pm  8.3$ & $270.0\pm 10.6$ & $-30.3\pm  3.5$\\
\\ \\
Q25TH000 & $1.621\pm0.020$ & $ 4.543 \pm 0.135$ & $ 5.521 \pm 0.044$ & $7.571\pm0.115$ & $145.1\pm 15.6$ & $855.3\pm 21.9$ & $535.3\pm  9.9$\\
Q25TH090 & $1.642\pm0.016$ & $-5.895 \pm 0.076$ & $ 4.412 \pm 0.039$ & $7.868\pm0.233$ & $-869.9\pm 19.8$ & $ 80.5\pm 14.2$ & $446.3\pm  7.4$\\
Q25TH130 & $1.530\pm0.012$ & $-6.879 \pm 0.040$ & $-0.096 \pm 0.005$ & $7.400\pm0.209$ & $-64.7\pm  9.1$ & $-141.3\pm  3.5$ & $-131.7\pm  2.2$\\
Q25TH210 & $1.638\pm0.018$ & $-1.045 \pm 0.106$ & $-7.132 \pm 0.021$ & $7.776\pm0.062$ & $-335.9\pm 27.8$ & $819.1\pm 17.2$ & $-487.8\pm  8.5$\\
Q25TH315 & $1.512\pm0.014$ & $ 6.774 \pm 0.063$ & $ 0.583 \pm 0.030$ & $7.375\pm0.188$ & $ 73.4\pm  7.1$ & $-45.9\pm  0.4$ & $223.4\pm  3.9$\\
\\ \\
Q33TH000 & $2.216\pm0.023$ & $ 4.731 \pm 0.073$ & $ 7.171 \pm 0.087$ & $11.516\pm0.098$ & $ 47.5\pm 19.1$ & $983.3\pm 19.6$ & $827.3\pm  8.8$\\
Q33TH090 & $2.184\pm0.018$ & $-7.569 \pm 0.026$ & $ 4.653 \pm 0.040$ & $11.548\pm0.082$ & $-829.1\pm 21.2$ & $-40.4\pm 15.5$ & $576.6\pm  8.0$\\
Q33TH130 & $2.050\pm0.016$ & $-8.485 \pm 0.017$ & $-0.840 \pm 0.053$ & $11.108\pm0.089$ & $171.0\pm  4.8$ & $  8.7\pm  0.2$ & $-374.3\pm  0.3$\\
Q33TH210 & $2.200\pm0.020$ & $-0.433 \pm 0.081$ & $-8.597 \pm 0.062$ & $11.515\pm0.032$ & $-453.0\pm 26.2$ & $793.0\pm 12.0$ & $-697.3\pm  6.7$\\
Q33TH315 & $2.054\pm0.017$ & $ 8.335 \pm 0.022$ & $ 1.495 \pm 0.010$ & $11.072\pm0.002$ & $306.9\pm  5.6$ & $155.7\pm  5.4$ & $520.5\pm  0.7$\\
\\ \\
Q40TH000 & $2.613\pm0.021$ & $ 4.604 \pm 0.114$ & $ 8.337 \pm 0.024$ & $14.791\pm0.074$ & $ 19.6\pm 19.4$ & $985.0\pm 15.0$ & $1132.5\pm  4.8$\\
Q40TH090 & $2.607\pm0.017$ & $-8.761 \pm 0.001$ & $ 4.729 \pm 0.133$ & $14.790\pm0.147$ & $-929.8\pm 18.2$ & $-83.4\pm 16.3$ & $939.3\pm  6.4$\\
Q40TH130 & $2.452\pm0.012$ & $-9.332 \pm 0.033$ & $-1.734 \pm 0.082$ & $14.123\pm0.130$ & $-12.3\pm  1.7$ & $-168.0\pm  5.5$ & $-137.0\pm  3.6$\\
Q40TH210 & $2.621\pm0.018$ & $ 0.157 \pm 0.047$ & $-9.604 \pm 0.013$ & $14.815\pm0.072$ & $-520.1\pm 24.9$ & $845.9\pm 10.4$ & $-1068.0\pm  5.8$\\
Q40TH315 & $2.441\pm0.014$ & $ 9.165 \pm 0.068$ & $ 2.464 \pm 0.083$ & $13.933\pm0.039$ & $120.6\pm  1.3$ & $-39.4\pm  0.6$ & $328.5\pm  2.9$\\
\\ \\
Q50TH000 & $2.814\pm0.013$ & $ 4.089 \pm0.0848$ & $ 7.762 \pm 0.092$ & $16.672\pm0.023$ & $-109.8\pm  3.1$ & $159.5\pm  7.5$ & $-77.3\pm  8.8$\\
Q50TH090 & $2.865\pm0.018$ & $-8.094 \pm0.0421$ & $ 4.566 \pm 0.198$ & $17.499\pm0.336$ & $623.4\pm  1.9$ & $-75.4\pm 13.3$ & $-1118.0\pm  7.8$\\
Q50TH130 & $2.991\pm0.017$ & $-9.082 \pm0.1534$ & $-2.150 \pm 0.025$ & $17.597\pm0.114$ & $688.2\pm  6.8$ & $611.5\pm 20.9$ & $-1330.0\pm  1.0$\\
Q50TH210 & $2.805\pm0.014$ & $ 0.087 \pm0.0538$ & $-8.922 \pm 0.098$ & $16.83\pm0.089$  & $  9.5\pm  4.9$ & $-221.3\pm  5.7$ & $590.5\pm  9.9$\\
Q50TH315 & $2.985\pm0.018$ & $ 8.788 \pm0.1391$ & $ 2.945 \pm0.0222$ & $17.43\pm0.0177$ & $600.4\pm  8.4$ & $665.1\pm 20.4$ & $1253.4\pm  2.4$\\
\\ \\
Q66TH000 & $3.277\pm0.012$ & $ 3.404 \pm0.1164$ & $ 7.633 \pm0.0533$ & $20.177\pm0.193$ & $-69.7\pm  0.9$ & $-294.6\pm  2.2$ & $-848.5\pm  9.2$\\
Q66TH090 & $3.419\pm0.013$ & $-8.028 \pm0.0443$ & $ 3.166 \pm0.2086$ & $21.293\pm0.133$ & $749.2\pm  6.2$ & $ 17.3\pm 11.4$ & $-1565.8\pm 11.9$\\
Q66TH130 & $3.344\pm0.012$ & $-7.896 \pm0.1482$ & $-2.756 \pm0.2232$ & $20.736\pm0.060$ & $356.3\pm  1.0$ & $425.9\pm 11.8$ & $-940.4\pm  2.3$\\
Q66TH210 & $3.336\pm0.013$ & $ 0.927 \pm0.1165$ & $-8.427 \pm0.0148$ & $20.701\pm0.048$ & $221.8\pm  7.5$ & $-487.5\pm  2.9$ & $1265.4\pm 11.0$\\
Q66TH315 & $3.321\pm0.011$ & $ 7.628 \pm0.1358$ & $ 3.368 \pm0.1800$ & $20.651\pm0.087$ & $259.2\pm  3.6$ & $404.5\pm 11.2$ & $750.7\pm  3.2$\\
\\ \\
Q100TH000& $3.580\pm0.006$ & $ 2.036 \pm0.1025$ & $ 6.224 \pm0.0869$ & $23.232\pm0.038$ & $-10.2\pm  2.6$ & $-447.0\pm  2.6$ & $-1354.5\pm  5.2$\\
Q100TH090& $3.538\pm0.004$ & $-6.452 \pm0.2779$ & $ 2.110 \pm0.2082$ & $23.541\pm0.332$ & $344.2\pm 11.2$ & $ 31.1\pm  3.3$ & $-962.9\pm  4.6$\\
Q100TH130& $3.461\pm0.001$ & $-6.393 \pm0.3509$ & $-2.042 \pm0.3155$ & $22.749\pm0.003$ & $-46.7\pm 14.3$ & $  6.8\pm  3.8$ & $245.1\pm 12.6$\\
Q100TH210& $3.606\pm0.011$ & $ 1.268 \pm0.0057$ & $-6.940 \pm0.5645$ & $23.382\pm0.072$ & $241.4\pm  0.6$ & $-418.6\pm  0.3$ & $1406.9\pm  2.8$\\
Q100TH315& $3.466\pm0.001$ & $ 6.268 \pm0.4289$ & $ 2.608 \pm0.2515$ & $22.748\pm0.035$ & $-77.6\pm 12.7$ & $-34.8\pm  3.4$ & $-383.3\pm 12.9$\\
\hline
Q25TH000HR & $1.605\pm0.022$& $4.623\pm0.025$& $5.699\pm0.014$& $7.503\pm0.124$& $110.2\pm  3.1$ & $539.2\pm  0.2$ & $471.9\pm  4.6$\\
Q25TH090HR & $1.632\pm0.022$& $-6.164\pm0.050$& $4.759\pm0.049$& $7.668\pm0.008$& $-698.5\pm  1.7$ & $104.1\pm  6.2$ & $504.7\pm  4.5$\\
Q25TH130HR & $1.736\pm0.021$& $-7.811\pm0.057$& $-0.682\pm0.012$& $7.793\pm0.012$& $-683.3\pm  2.4$ & $-549.1\pm 13.4$ & $400.6\pm  1.6$\\
Q25TH210HR & $1.629\pm0.022$& $-1.129\pm0.002$& $-7.430\pm0.038$& $7.622\pm0.134$& $-245.6\pm  8.2$ & $656.9\pm  0.9$ & $-512.7\pm  4.4$\\
Q25TH315HR & $1.715\pm0.020$& $7.618\pm0.043$& $1.323\pm0.023$& $7.692\pm0.005$& $-539.0\pm  1.7$ & $-549.2\pm 11.6$ & $-300.4\pm  0.7$\\
\end{tabular}
\end{ruledtabular}
\end{table}
\end{widetext}

In order to measure the dependence of our recoil calculations on the
choice of $r_{+}$, $r_{0}$, $r_{-}$ (the three points in the
trajectory that define the late-time orbital plane) we make two
choices for these quantities a fiducial choice $(r_{+}, r_{0}, r_{-})/M = (2.0,
1.0, 0.5)$ and a second choice $(r_{+}, r_{0}, r_{-})/M = 2.2, 1.2, 0.7)$
based on the point in the orbital trajectory when $|\dot r(r)|$ is
a maximum.  For our runs we found that the maximum in
$|\dot r(t)|$ occurs at $r=(1.2\pm0.1)$. 
We also use an alternative approach, discussed in Sec.~\ref{sec:alt_plane}, where
we choose $(r_{+}, r_{0}, r_{-})$ to be the extrema of $\dot r(t)$ and $\ddot r(t)$ during the
plunge phase. In this latter approach, the values of $(r_{+}, r_{0}, r_{-})$ vary with
each configuration and generally produce poorer results than for the fixed choice
of $(r_{+}, r_{0}, r_{-})$.
In
Table~\ref{table:transformedkick} we show how these two choices affect
the calculation of $v_{\|}$ and $v_\perp$, the spin component in and
out of the plane, and the angle between the in-plane spins with the
corresponding spin for the TH000 configurations ($\vartheta$). 
Note that we report the
magnitude of the in-plane (i.e.\ $v_\perp$) recoil and spin $a_\perp$, because modifying
$(r_{+}, r_{0}, r_{-})$ introduces additional rotations within the orbital
plane; making a direct comparison of $x$ and $y$ components meaningless.
Note that there is scatter in both the in-plane and out-plane
components of the spin. The scatter in the out-of-plane component
appears to be relatively significant due to its small average value. The source
of this scatter may simply be due to  errors in estimating the true direction
of the orbital plane (since the spin in the plane is large). We take
the average value of the in-plane spin when fitting to
Eq.~(\ref{eq:empirical}) below.
\begin{widetext}

\begin{table}
\caption{The transformed spin (at $r=r_0$), recoil velocities in $\KMS$, and
angles between the in-plane QXXXTH000 spins and the other QXXXTHYYY
configurations in degrees. A `$1$' denotes a quantity measured using  $(r_{+},
r_{0}, r_{-})/M = (2.0,
1.0, 0.5)$, while a `$2$' denotes a quantity measured using    $(r_{+},
r_{0}, r_{-})/M =(2.2, 1.2,0.7)$. The differences between the values
given for $(2.0,
1.0, 0.5)$ and $(2.2, 1.2,0.7)$ are indicative of the errors. Note
that in the transformed system $L_z$ and $S_z$ are both negative
(i.e.\ there is some partial spin/orbit alignment. We denote
quantities in the (transformed) orbital plane with
a $\perp$ subscript. The Q25THXXX quantities given on the bottom of
the table are for the higher resolution runs. See
Sec.~\ref{sec:finite_diff} for an explanation of the differences.
The time derivative of the orbital separation $|\dot r|$ is a maximum at
$r\sim1.2$.}
  \label{table:transformedkick}
\begin{ruledtabular}
\begin{tabular}{lcccccccccc}
Config   & $V_{\perp} (1)$ & $V_{\|} (1)$ & $a_{\perp}/m^H (1)$ & $a_{\|}/m^H (1)$ & $\vartheta (1)$ & $V_{\perp} (2)$ & $V_{\|} (2)$ & $a_{\perp}/m^H (2)$ & $a_{\|}/m^H (2)$ & $\vartheta (2)$ \\ \hline
Q13TH000 & 57.1362 & -109.88 & 0.768518 & -0.237098 & 0. &  49.953 & -113.327 & 0.751038 & -0.291919 & 0. \\
Q13TH090 & 64.941 & -97.8627 & 0.773066 & -0.22036 & 3.63707 &  58.4083 & -101.897 & 0.756446 & -0.277112 & 3.96937 \\
Q13TH130 & 220.34 & -368.446 & 0.727118 & -0.332954 & -78.283 &  249.903 & -349.071 & 0.761071 & -0.249773 & -77.2282 \\
Q13TH210 & 42.1564 & 128.679 & 0.763202 & -0.24517 & 177.413 &  31.9594 & 131.582 & 0.746572 & -0.295965 & 177.276 \\
Q13TH315 & 239.084 & 369.106 & 0.742796 & -0.295281 & 87.2139 &  280.513 & 338.693 & 0.772431 & -0.212227 & 89.2203 \\
\\
Q17TH000 & 309.622 & 568.413 & 0.751456 & -0.270601 & 0. &  370.742 & 530.574 & 0.776806 & -0.193091 & 0. \\
Q17TH090 & 134.116 & 344.786 & 0.735493 & -0.317919 & 67.034 &  123.344 & 348.784 & 0.738562 & -0.314279 & 63.9956 \\
Q17TH130 & 351.795 & 366.112 & 0.78253 & -0.15129 & -36.082 &  393.985 & 320.27 & 0.786039 & -0.145002 & -35.7209 \\
Q17TH210 & 300.137 & -563.416 & 0.746093 & -0.279011 & -176.311 &  357.269 & -529.036 & 0.773073 & -0.199299 & -176.459 \\
Q17TH315 & 283. & -156.073 & 0.784854 & -0.145801 & 123.407 &  295.546 & -130.769 & 0.778357 & -0.186578 & 123.21 \\
\\
Q25TH000 & 373.905 & -948.305 & 0.757314 & -0.240225 & 0. &  459.51 & -909.911 & 0.77861 & -0.167965 & 0. \\
Q25TH090 & 452.721 & -870.357 & 0.779284 & -0.154404 & -21.2857 &  544.717 & -815.942 & 0.788089 & -0.114385 & -21.0555 \\
Q25TH130 & 197.85 & 48.3342 & 0.770647 & -0.202858 & -84.2701 &  195.719 & 56.3483 & 0.759915 & -0.247894 & -85.4253 \\
Q25TH210 & 426.136 & 916.571 & 0.77451 & -0.184646 & 165.452 &  521.338 & 865.968 & 0.787368 & -0.131325 & 165.72 \\
Q25TH315 & 124.649 & -204.556 & 0.759673 & -0.241429 & 85.9471 &  115.438 & -209.892 & 0.750901 & -0.274232 & 84.2327 \\
\\
Q33TH000 & 477.923 & -1193.82 & 0.787385 & -0.136406 & 0. &  585.012 & -1145.16 & 0.794072 & -0.101676 & 0. \\
Q33TH090 & 506.036 & -874.874 & 0.794658 & -0.0857209 & -23.2403 &  573.118 & -832.474 & 0.794033 & -0.100666 & -23.2941 \\
Q33TH130 & 121.993 & 393.11 & 0.754769 & -0.264995 & -83.3346 &  106.146 & 397.682 & 0.753714 & -0.275061 & -85.3148 \\
Q33TH210 & 510.834 & 1029.22 & 0.793461 & -0.100511 & 165.757 &  597.049 & 981.726 & 0.795102 & -0.0964032 & 165.86 \\
Q33TH315 & 142.073 & -607.563 & 0.74435 & -0.292739 & 85.1084 &  124.37 & -611.433 & 0.751109 & -0.280681 & 82.8364 \\
\\
Q40TH000 & 395.545 & -1447.96 & 0.782576 & -0.177756 & 0. &  501.737 & -1414.67 & 0.792558 & -0.121423 & 0. \\
Q40TH090 & 506.998 & -1223.4 & 0.797605 & -0.0811617 & -29.708 &  599.942 & -1180.61 & 0.798469 & -0.074772 & -29.6255 \\
Q40TH130 & 199.308 & 85.9742 & 0.773172 & -0.20708 & -88.1121 &  196.752 & 91.6738 & 0.767006 & -0.237119 & -89.0041 \\
Q40TH210 & 468.966 & 1380.9 & 0.794067 & -0.117055 & 163.468 &  578.675 & 1338.63 & 0.797642 & -0.0855529 & 163.572 \\
Q40TH315 & 138.257 & -323.823 & 0.762353 & -0.243806 & 82.1551 &  127.284 & -328.291 & 0.760328 & -0.257681 & 80.7494 \\
\\
Q50TH000 & 206.386 & 29.5953 & 0.775871 & -0.197482 & 0. &  205.713 & 33.9596 & 0.771006 & -0.223791 & 0. \\
Q50TH090 & 211.758 & 1264.66 & 0.760737 & -0.256772 & -53.0652 &  228.683 & 1261.71 & 0.77738 & -0.20449 & -53.2402 \\
Q50TH130 & 426.349 & 1560.33 & 0.793283 & -0.105245 & -109.529 &  530.071 & 1528.21 & 0.800072 & -0.0682939 & -108.518 \\
Q50TH210 & 117.707 & -619.58 & 0.759212 & -0.259 & 158.194 &  100.605 & -622.586 & 0.763761 & -0.251091 & 157.494 \\
Q50TH315 & 454.34 & -1472.23 & 0.799245 & -0.0784754 & 62.2872 &  549.566 & -1439.4 & 0.799962 & -0.065415 & 63.1927 \\
\\
Q66TH000 & 118.274 & 893.123 & 0.763607 & -0.251388 & 0. &  101.713 & 895.16 & 0.772394 & -0.226846 & 0. \\
Q66TH090 & 268.896 & 1714.94 & 0.7931 & -0.116783 & -62.1364 &  354.418 & 1699.33 & 0.802482 & -0.066631 & -60.9979 \\
Q66TH130 & 336.909 & 1038.86 & 0.80055 & -0.0605261 & -111.577 &  377.096 & 1024.96 & 0.797946 & -0.0868091 & -110.371 \\
Q66TH210 & 163.971 & -1364.27 & 0.770063 & -0.232589 & 157.52 &  176.949 & -1362.65 & 0.783732 & -0.184392 & 157.808 \\
Q66TH315 & 311.388 & -835.099 & 0.798347 & -0.0778683 & 60.8859 &  339.961 & -823.881 & 0.795121 & -0.109404 & 62.06 \\
\\
Q100TH000 & 88.5695 & 1423.66 & 0.79471 & -0.158448 & 0. &  108.207 & 1422.3 & 0.800356 & -0.116206 & 0. \\
Q100TH090 & 134.476 & 1014.12 & 0.806191 & -0.0286805 & -65.6896 &  167.057 & 1009.26 & 0.803238 & -0.0439484 & -65.4866 \\
Q100TH130 & 52.4647 & -244.074 & 0.785337 & -0.163234 & -116.365 &  42.6164 & -245.985 & 0.785246 & -0.170854 & -116.398 \\
Q100TH210 & 95.9383 & -1484.52 & 0.805662 & -0.0736974 & 151.702 &  151.312 & -1479.91 & 0.804525 & -0.0475139 & 151.917 \\
Q100TH315 & 55.0809 & 388.766 & 0.78288 & -0.178353 & 58.0786 &  40.6688 & 390.537 & 0.78611 & -0.178897 & 57.8702 \\
\hline
Q25TH000HR & 238.923 & -684.464 & 0.727703 & -0.328834 & 0 & 235.693 & -685.583 & 0.744828 & -0.292768 & 0 \\
Q25TH090HR & 302.537 & -813.638 & 0.731821 & -0.318884 & -12.0909 & 317.701 & -807.838 & 0.754461 & -0.263398 & -11.4989 \\
Q25TH130HR & 514.081 & -815.226 & 0.787921 & -0.127218 & -83.1554 & 596.917 & -756.679 & 0.791774 & -0.114032 & -80.3582 \\
Q25TH210HR & 299.288 & 815.542 & 0.730048 & -0.325618 & 167.319 & 315.17 & 809.537 & 0.753828 & -0.269311 & 167.829 \\
Q25TH315HR & 501.346 & 656.57 & 0.79009 & -0.113807 & 85.1511 & 560.613 & 606.749 & 0.789798 & -0.127652 & 87.8788 \\
\end{tabular}
\end{ruledtabular}
\end{table}
\end{widetext}

\begin{widetext}

\begin{figure}
\caption{A fit of $v_\|$ for the $q=1, 2/3, 1/2, 1/2.5, 1/3, 1/4,
1/6, 1/8$, and the high-resolution $q=1/4$ runs (from
top to bottom) with the plane chosen by the points on the trajectory
with $(r_{+}, r_{0}, r_{-})=(2,1,0.5)$ (left)
 and $(r_{+}, r_{0}, r_{-})=(2.2,1.2,0.7)$ (right).}
\includegraphics[width=2.5in,height=1.0in]{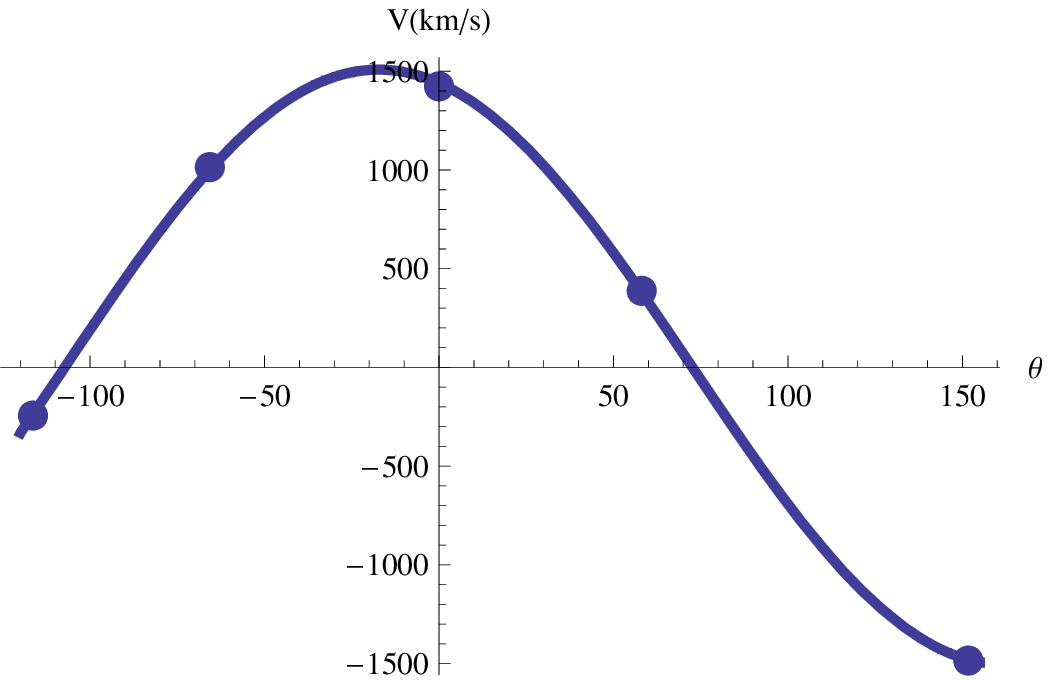}
\includegraphics[width=2.5in,height=1.0in]{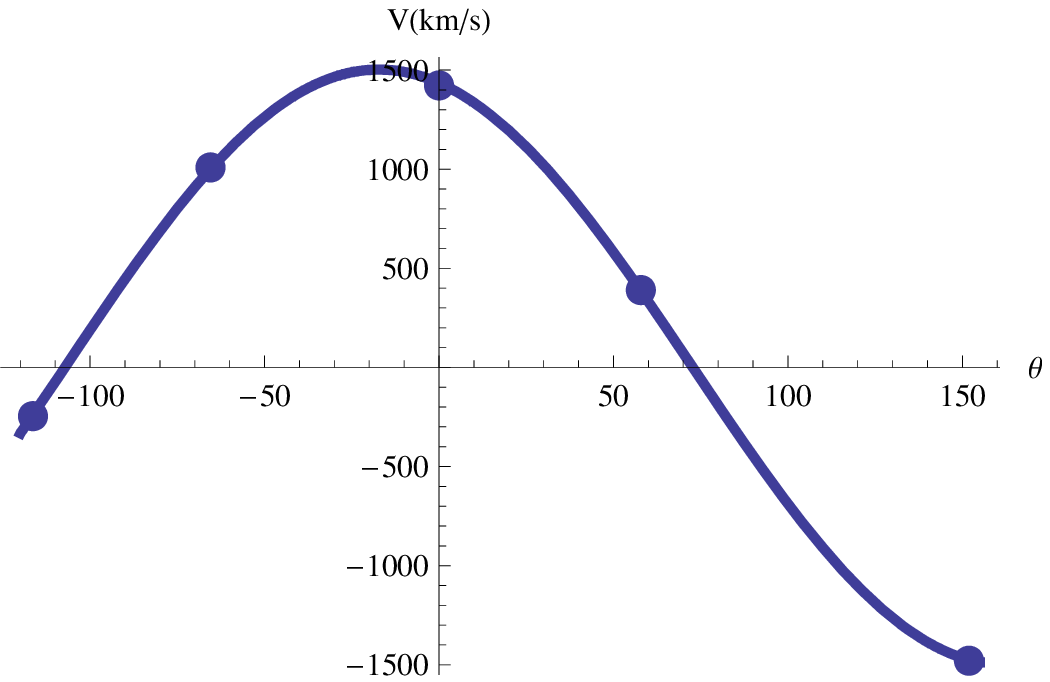}
\includegraphics[width=2.5in,height=1.0in]{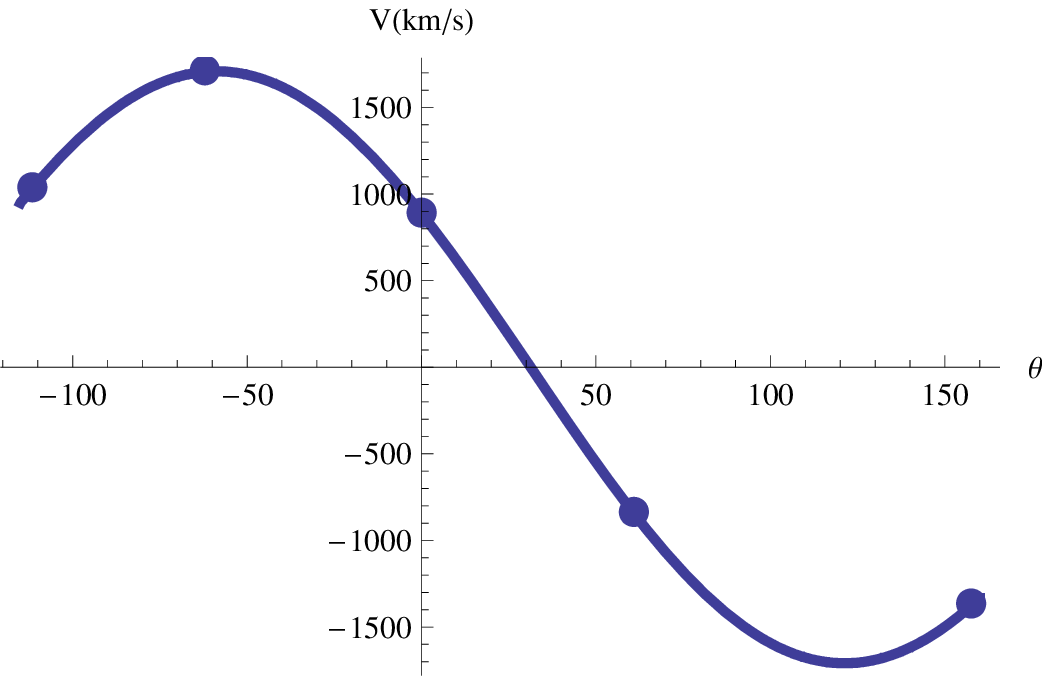}
\includegraphics[width=2.5in,height=1.0in]{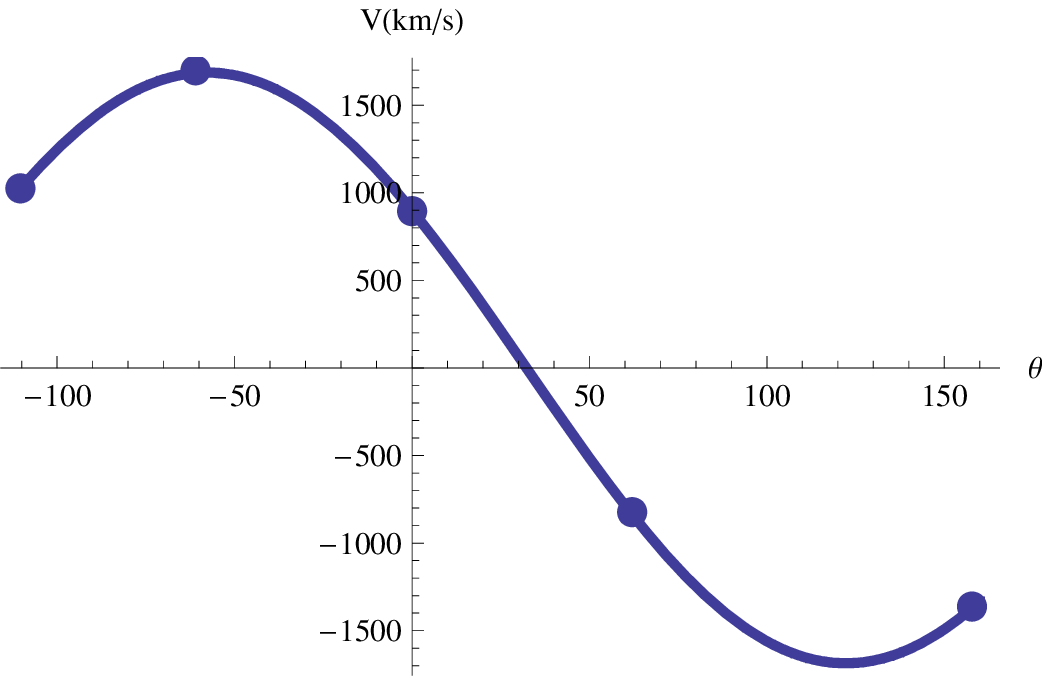}
\includegraphics[width=2.5in,height=1.0in]{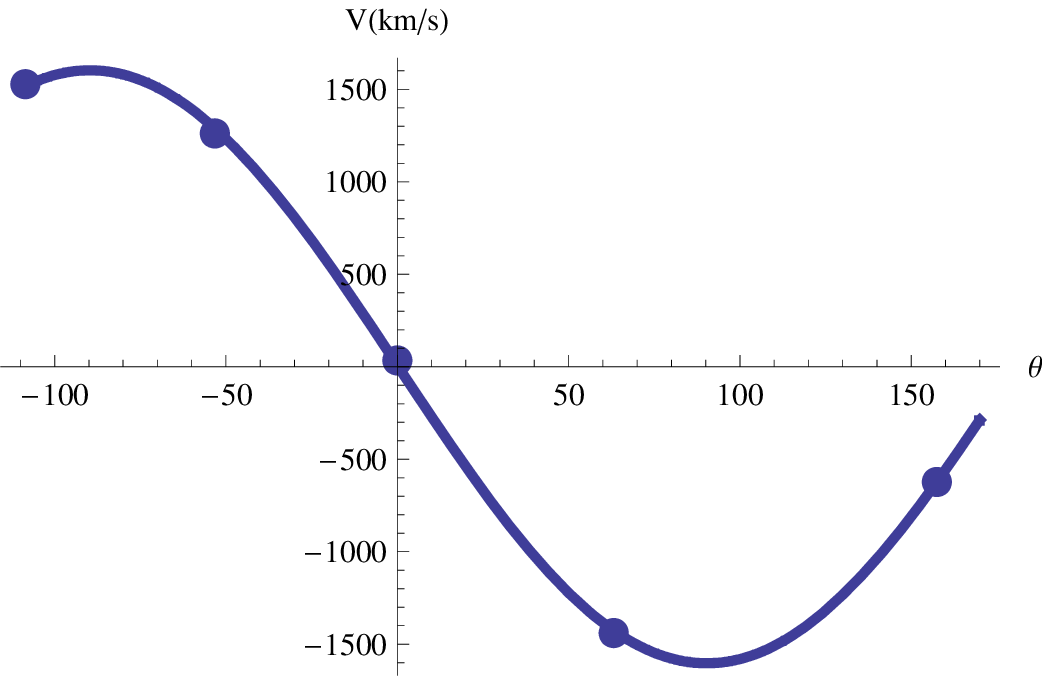}
\includegraphics[width=2.5in,height=1.0in]{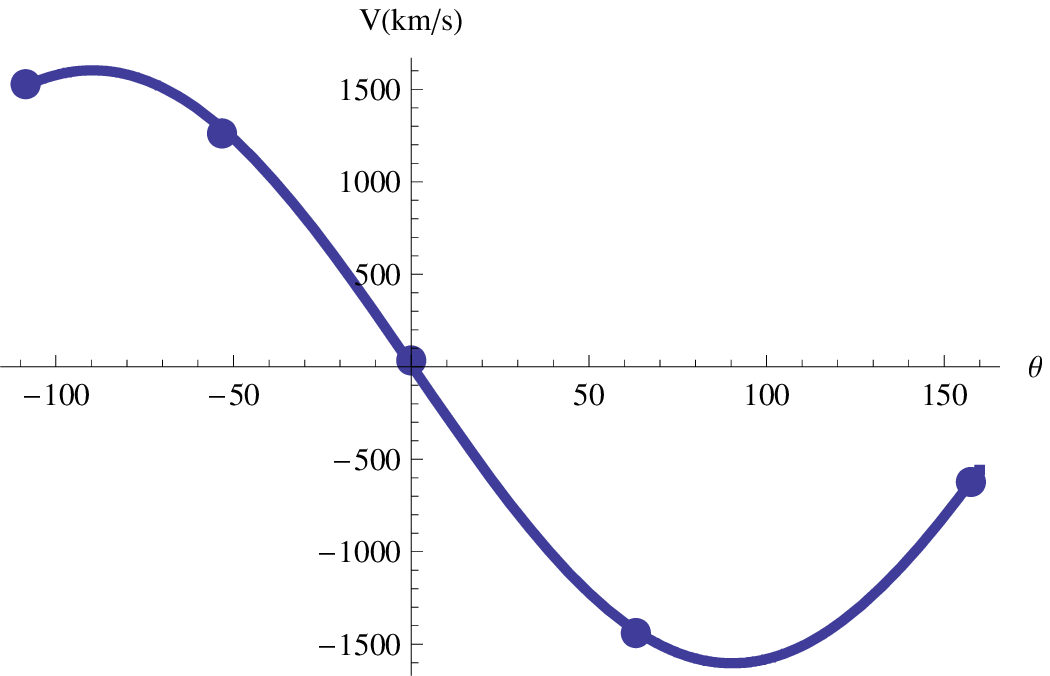}
\includegraphics[width=2.5in,height=1.0in]{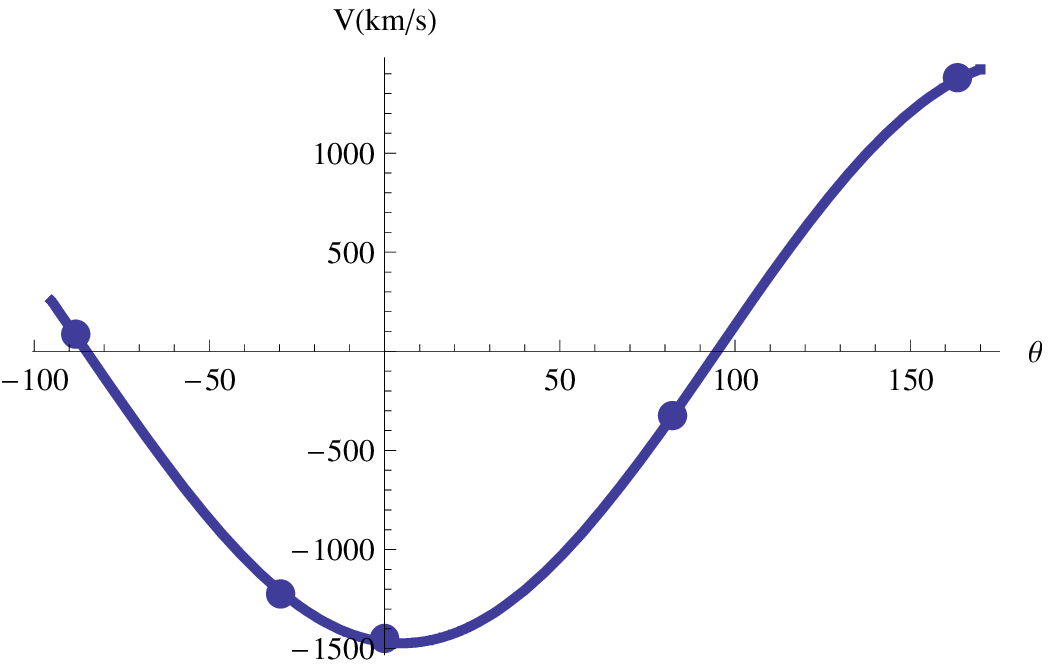}
\includegraphics[width=2.5in,height=1.0in]{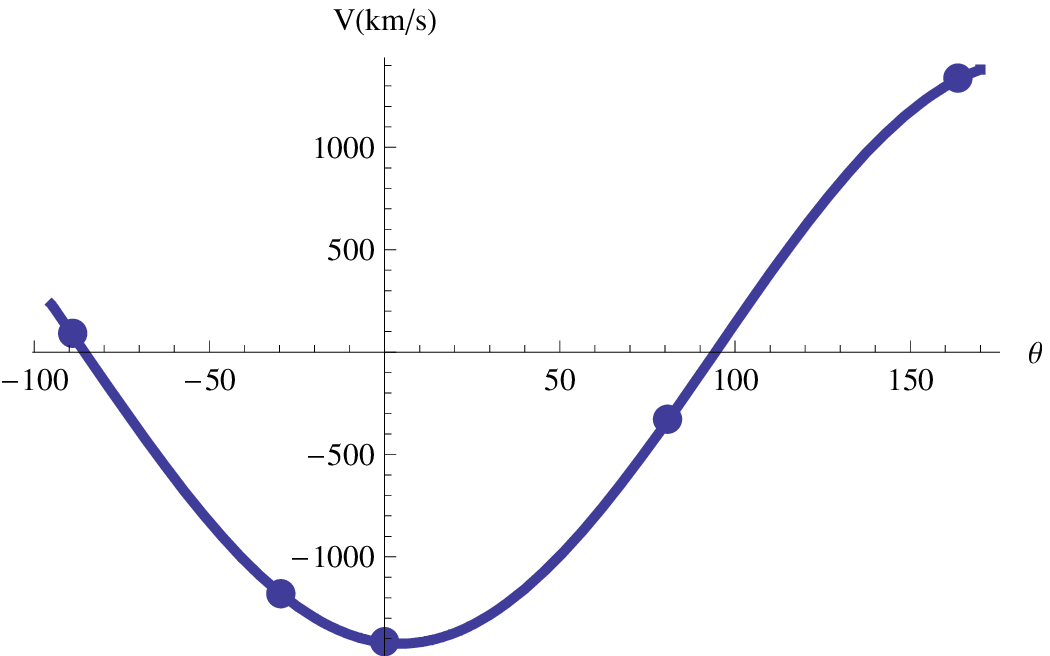}
\includegraphics[width=2.5in,height=1.0in]{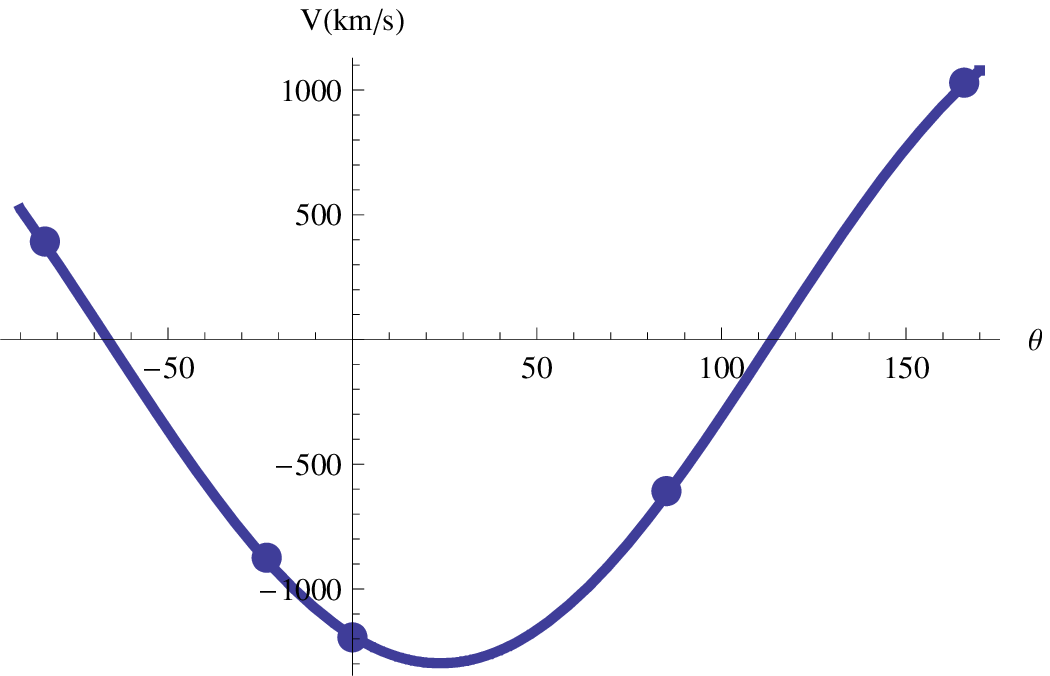}
\includegraphics[width=2.5in,height=1.0in]{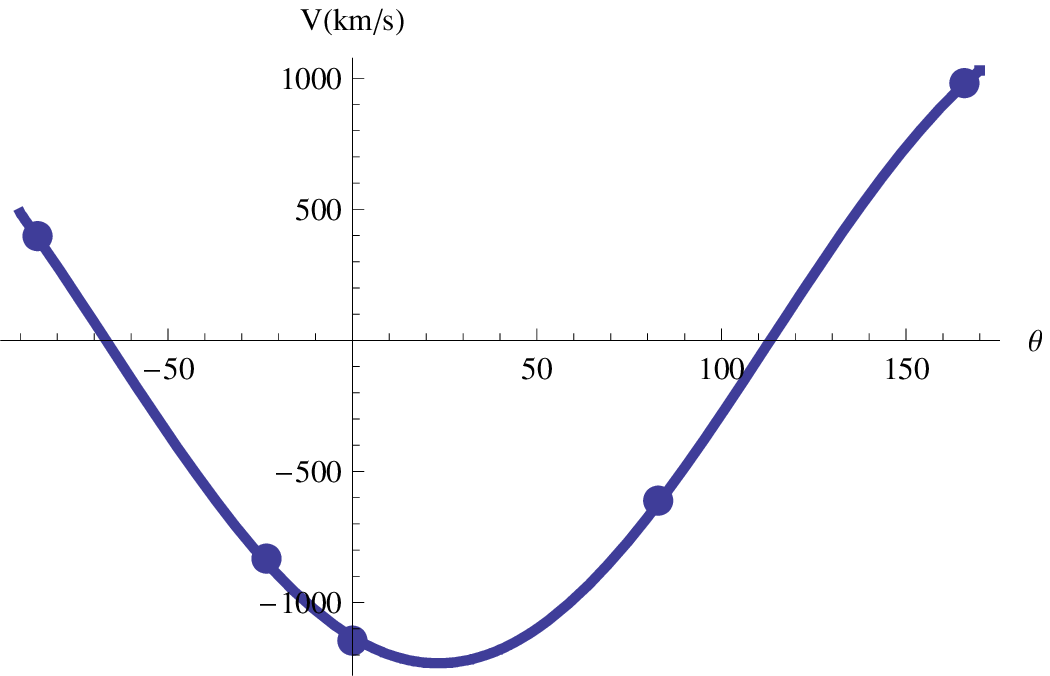}
\includegraphics[width=2.5in,height=1.0in]{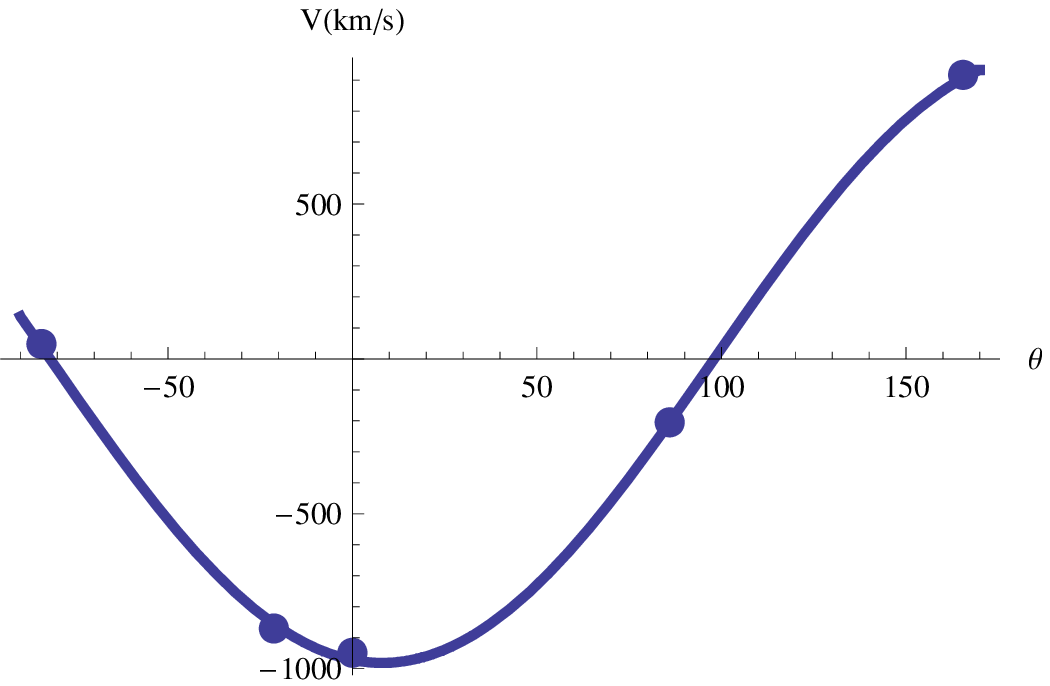}
\includegraphics[width=2.5in,height=1.0in]{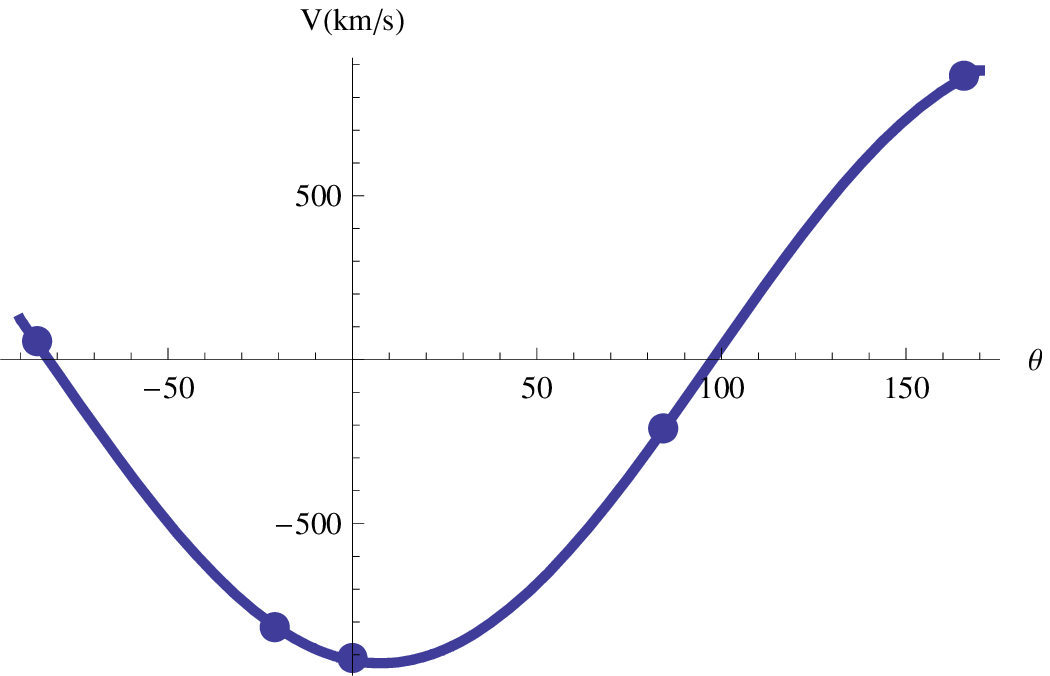}
\includegraphics[width=2.5in,height=1.0in]{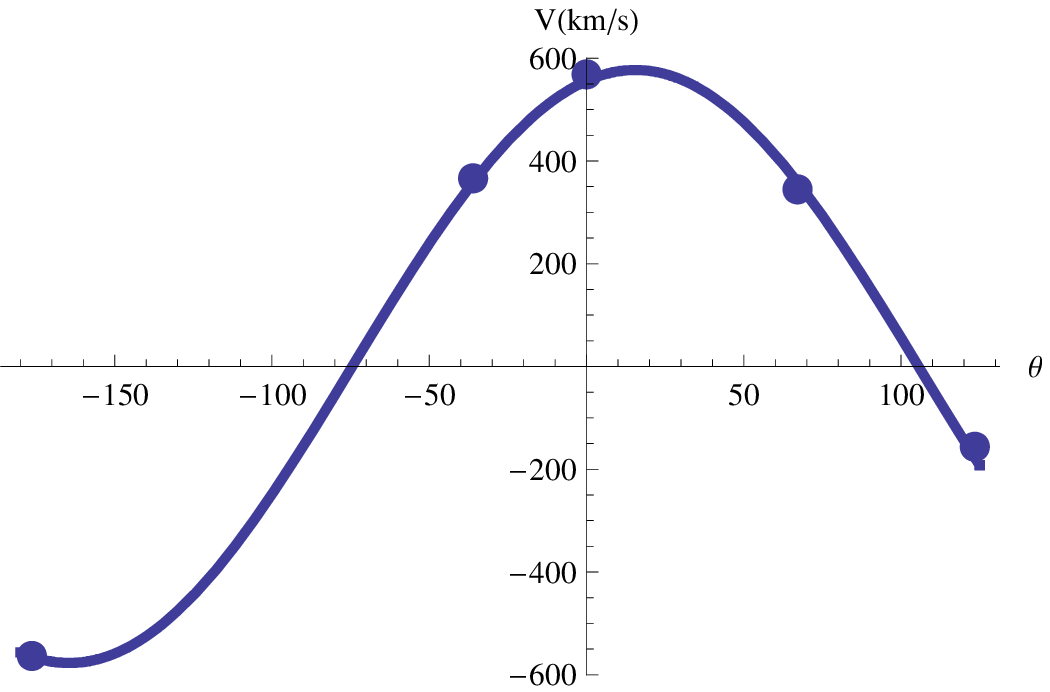}
\includegraphics[width=2.5in,height=1.0in]{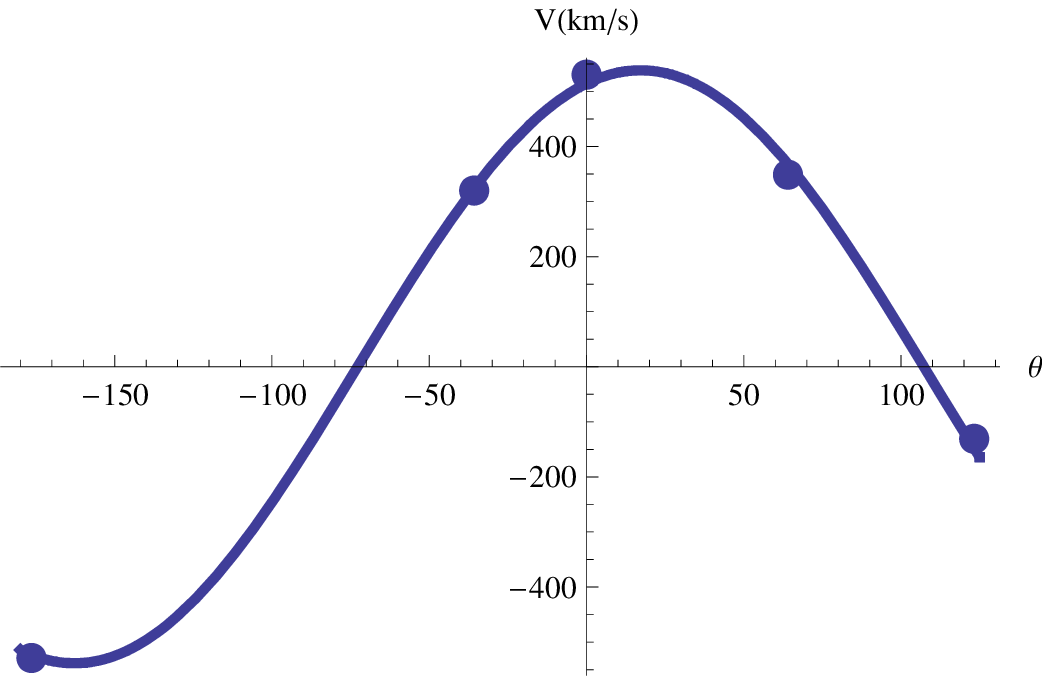}
\includegraphics[width=2.5in,height=1.0in]{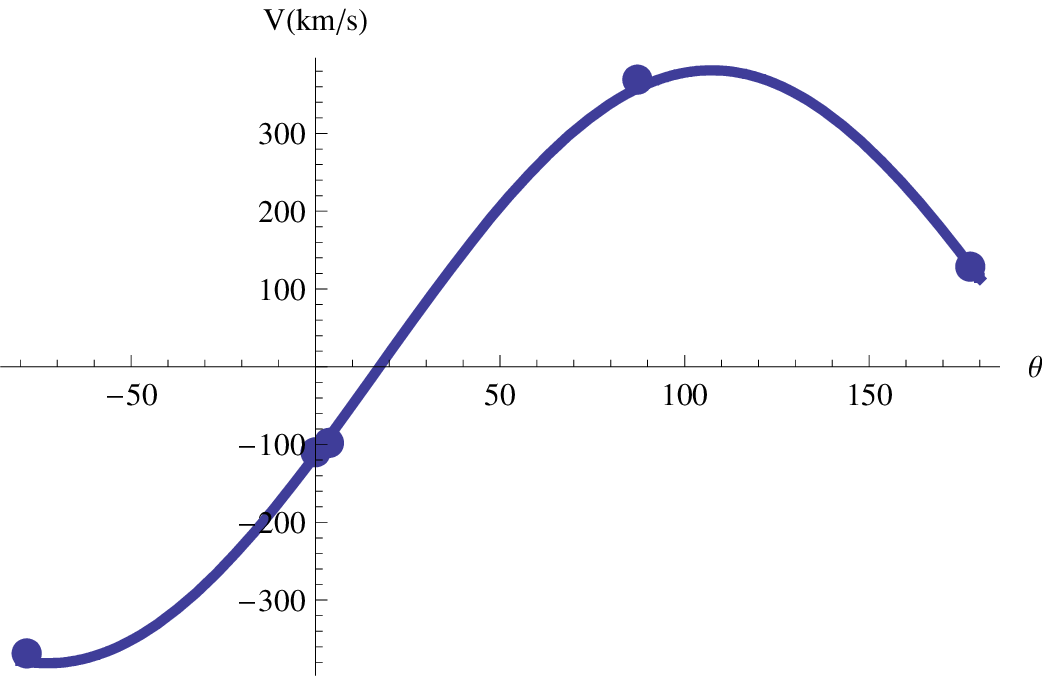}
\includegraphics[width=2.5in,height=1.0in]{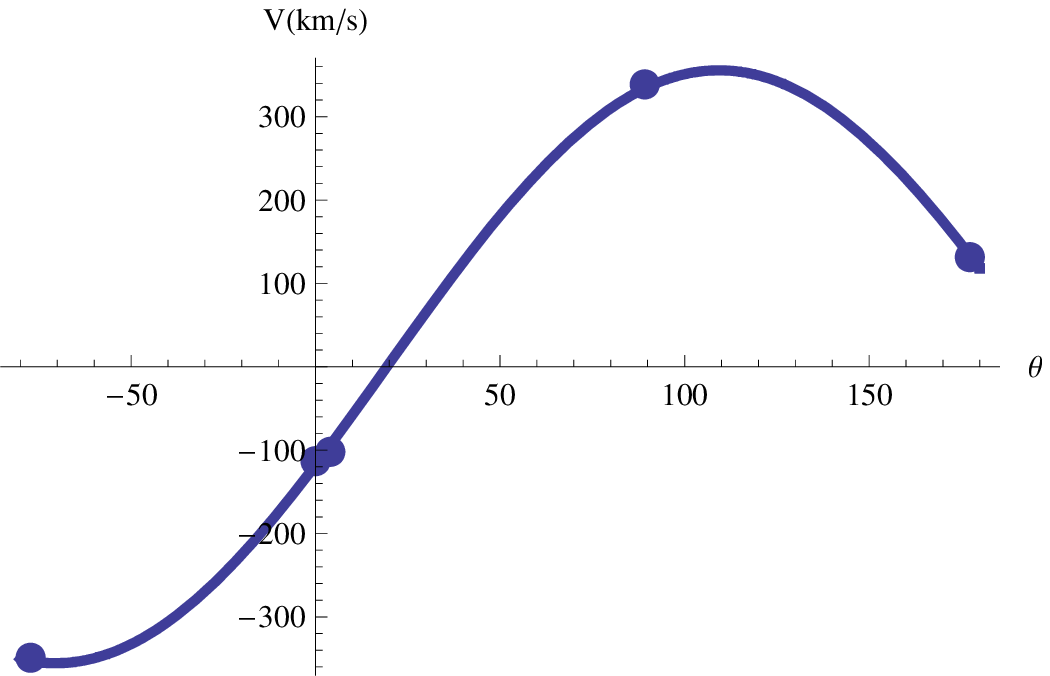}
\includegraphics[width=2.5in,height=1.0in]{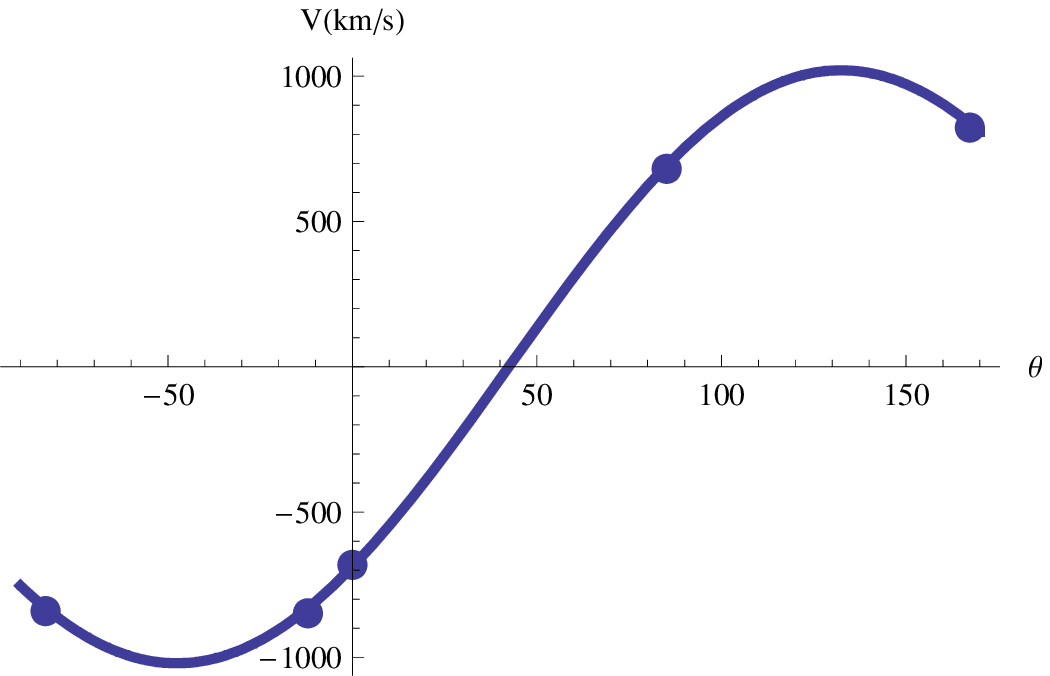}
\includegraphics[width=2.5in,height=1.0in]{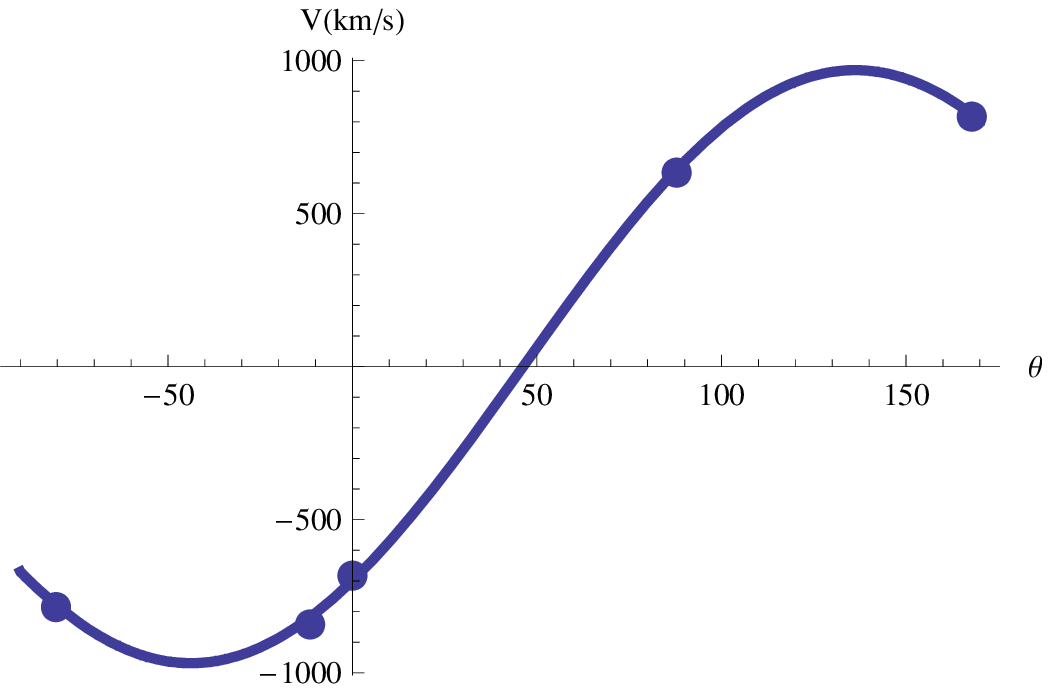}
\label{fig:theta_fits}
\end{figure}
\end{widetext}

Our empirical formula~(\ref{eq:empirical}) predicts that $v_\|$ will
scale as $\cos (\vartheta - \vartheta_0)$. For each set of
configurations with a given mass ratio, we perform a non-linear least
squares fit of $v_\|$ to the form 
\begin{equation}
v_\| = A \cos(\vartheta -B)
\label{eq:v_v_th}
\end{equation}
 and
solve for $A$ and $B$. The results of these fits are summarized in
Table~\ref{table:vofq}. 
\begin{table}
\caption{Fit parameters for $v_\| = A \cos((\vartheta - B)\pi/180)$ for each $q$.
$A$ is in units of $\KMS$ and $B$ is in degrees. A (1) denotes values obtained from the
$(r_{+}, r_{0}, r_{-})=(2,1,0.5)$ transformation, while a (2) denotes values obtained from the
$(r_{+}, r_{0}, r_{-})=(2.2,1.2,0.7)$ transformation. The last row
shows $A$ and $B$ when the grid is refined by a factor of $1.2$}
\label{table:vofq}
\begin{ruledtabular}
\begin{tabular}{lcccc}
$q$ & A(1) & B(1) & A(2) & B(2) \\
\hline
1/8 & $381.2\pm7.1$ & $107.33\pm0.93$    & $355.7\pm4.4$ & $109.35\pm0.63$\\
1/6 & $577\pm11$ & $15.51\pm1.18$        & $538\pm12$ & $17.2\pm1.4$ \\
1/4 & $981\pm11$ & $188.03\pm0.70$       & $926.3\pm5.1$ & $187.61\pm0.33$ \\
1/3 & $1297.8\pm9.2$ & $203.75\pm0.37$   & $1231.7\pm9.5$ & $203.22\pm0.41$ \\
1/2.5 & $1472.4\pm9.6$ & $184.88\pm0.39$ & $1424.1\pm5.1$ & $184.35\pm0.21$ \\
1/2 & $1640\pm23$ & $269.96\pm0.77$      & $1603\pm15$ & $270.32\pm0.54$ \\
2/3 & $1709.1\pm7.7$ & $301.26\pm0.26$   & $1686.0\pm8.6$ & $246.04\pm0.30$\\
1 & $1508.1\pm8.7$ & $342.69\pm0.32$     & $1502.5\pm6.7$ & $342.76\pm0.25$ \\
\hline
1/4 & $994.6\pm 6.4$ & $133.03\pm 0.43$ & $942.5\pm4.6$ & $137.09\pm0.35$\\
\end{tabular}
\end{ruledtabular}
\end{table}
In Fig.~\ref{fig:theta_fits} we plot the individual
data points for each mass ratio and the best fit function.
Note that the effect of refining the grid is to slightly increase the 
magnitude of the out-of-plane recoil.

\begin{figure}
\caption{Fit of out-of-plane recoil ($V_\|/\alpha$) versus mass ratio
$q$ for the $(r_{+}, r_{0}, r_{-})=(2,1,0.5)$ choice of orbital plane.
The top shaded region shows our predicted value for
$K=(6.0\pm0.2)\times 10^4$, the lower region shows the prediction
based on the modification proposed by Baker et al., the lower (green)
curve shows a fit to $\eta^n$, while the upper (blue) curve has a
simultaneous fit to $\eta^n$ and $K$. Note that the measured values of
$V_\|/\alpha$, while slightly overshooting the predictions, agree much
better with our ${\cal O}(\eta^2)$ form. The dots are the locations of
the data points.}
\includegraphics[width=3.0in]{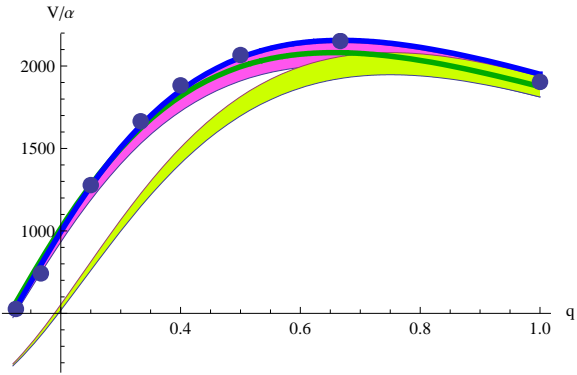}
\label{fig:r2105_fit}
\end{figure}

\begin{figure}
\caption{Fit of out-of-plane recoil ($V_\|/\alpha$) versus mass ratio $q$
for the $(r_{+}, r_{0}, r_{-})=(2.2,1.2,0.7)$ choice of orbital plane.
The top shaded region shows our predicted value for
$K=(6.0\pm0.2)\times 10^4$, the lower region shows the prediction
based on the modification proposed by Baker et al., the lower (green)
curve shows a fit to $\eta^n$, while the upper (blue) curve has a
simultaneous fit to $\eta^n$ and $K$. Note that the measured values of
$V_\|/\alpha$, while slightly overshooting the predictions, agree much
better with our ${\cal O}(\eta^2)$ form. The dots are the locations of
the data points.}
\includegraphics[width=3.0in]{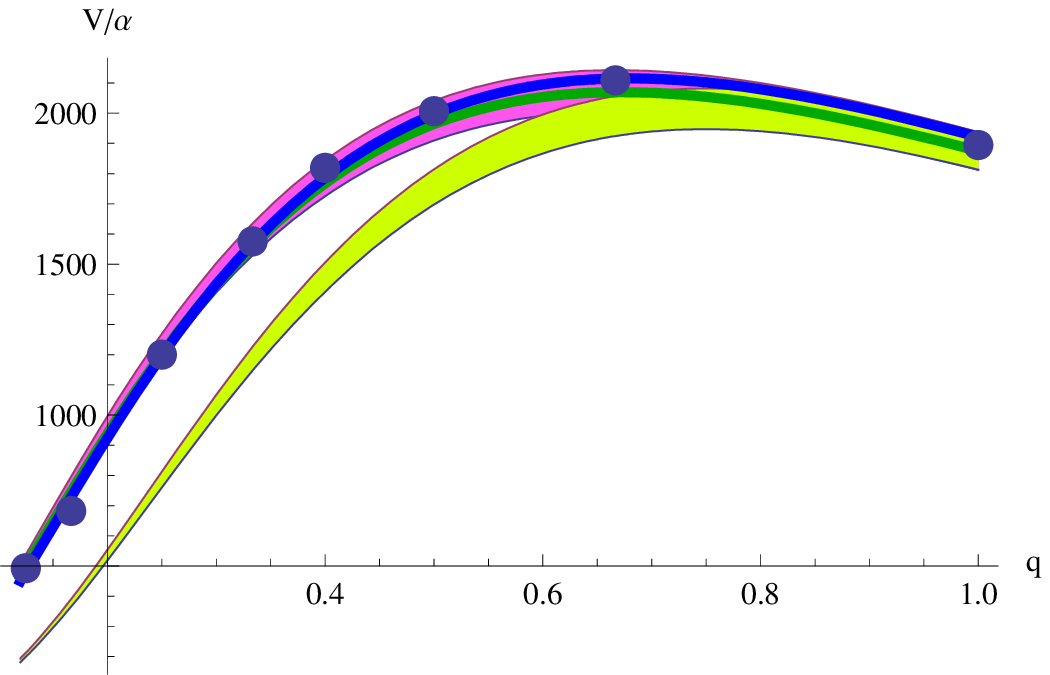}
\label{fig:r2.21.207_fit}
\end{figure}

\begin{figure}
\caption{The out-of-plane recoil ($V_\|/\alpha$) versus mass ratio $q$
for both choices of $(r_{+}, r_{0}, r_{-})$ as well as the predictions
assuming a leading $\eta^2$ and $\eta^3$ dependence. The solid lines
show the prediction assuming a leading $\eta^2$ dependence and an error
in $K$ of $0.2\times 10^4\ \KMS$, while the dotted lines show the
prediction assuming a leading $\eta^3$ dependence and an error
in $K$ of $0.2\times 10^4\ \KMS$. The $((r_{+}, r_{0}, r_{-}) =
(2,1,0.5)$ data points lie above the $(2.2,1.2.0.7)$ data points.
The $(2.2,1.2.0.7)$ data points lie closer to the predicted values, which
is consistent with the observation that $r_0 =1.2$ is a better approximation
to the location of the maximum of $|\dot r|$ for all configurations than
$|\dot r=1|$. 
The differences in these data points (for a given $q$) is indicative
of the true error in calculating the out-of-plane recoil.}
\includegraphics[width=3.0in]{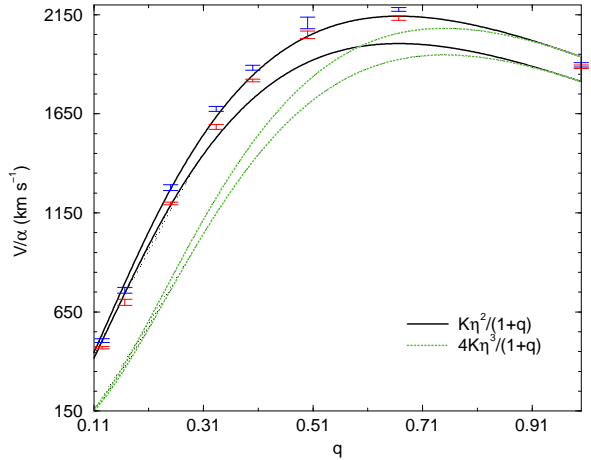}
\label{fig:fit}
\end{figure}

With the above results we can now fit the maximum out-of-plane recoil
($v_\|$) versus $q$. Our empirical formula~(\ref{eq:empirical})
predicts that this maximum will have the form
 $v_\| = K \eta^2/(1+q) (\alpha_1 - q \alpha_2)$, where $\eta = q /
(1+q)^2$. For all of our configurations $\alpha_2 \approx 0$, and hence
$v_\| = \alpha_1 K \eta^2/(1+q)$. Interestingly, this form has a
maximum at $q=2/3$. Baker et al.~\cite{Baker:2008md} propose that
Eq.~(\ref{eq:empirical}) should be modified to $v_\| = 4 K \eta^3 /
(1+q) (\alpha_1 - q \alpha_2)$ (which has a maximum at $q=3/4$ for
fixed $\alpha_1$ and $\alpha_2 =0$). In order to discriminate between these
two possibilities we fit our data to the form
\begin{equation}
  \label{eq:ffit}
  v_\|/\alpha = G (4 \eta)^H/(16(1+q)),
\end{equation}
where  we set
 $G=K$ and  solve for $H$, as well as allowing
both $G$ and $H$ to vary. Here $\alpha = |\alpha_1^\perp -
 q \alpha_2^\perp|$. It is important to note that our fits use
eight values of $q$ to obtain one or two parameters. In
Figs.~\ref{fig:r2105_fit}, \ref{fig:r2.21.207_fit}, and \ref{fig:fit}
 we show results
from these fits for the choices $(r_{+}, r_{0}, r_{-})=(2,1,0.5)$ and
$(r_{+}, r_{0}, r_{-})=(2.2,1.2,0.7)$.
In the figures we plot the predicted recoil velocity using both our
formula and the one proposed by Baker et al. assuming an uncertainty in
$K$ twice as large as that given in~\cite{Campanelli:2007cga}. From
the plots we can see that the recoil velocities agree with our
empirical formula much better than with the Baker et al. modification.
The best fit functions have the form $v_\|/\alpha = K \eta^{1.91\pm0.062} /
(1+q)$ (with a root mean square error in the predicted $v/\alpha$ of
$68\ \KMS$) and $v_\|/\alpha = K \eta^{2.036\pm0.046} /
(1+q)$ (with a root mean square error in $v/\alpha$ of
$46\ \KMS$) for the choices $(r_{+}, r_{0}, r_{-})=(2,1,0.5)$ and $(r_{+},
r_{0}, r_{-})=(2.2,1.2,0.7)$
respectively, where $K$ was set to $K=6.0\times 10^4$, and
$v_\|/\alpha =
(62965\pm 746) \eta^{2.027\pm0.048} / (1+q)$ (with a root mean square
error in the predicted $v/\alpha$ of $35\ \KMS$) and $v_\|/\alpha =
(62024\pm555) \eta^{2.127\pm0.037} / (1+q)$ (with a root mean square
error in the predicted $v/\alpha$ of $26\ \KMS$) respectively when both
$G$ and $H$ are varied. 
The value of the constant $K=G_{\rm fit}$ determined in these latter two fits is
in reasonable agreement with our previous measurement of
$K=(6.0\pm0.1)\times 10^4$.

\subsection {Finite difference errors}
\label{sec:finite_diff}
The effect of finite difference errors are most significant for the
smaller $q$ runs. Here the main effect regarding our calculation 
is that the $\vartheta$ is
rotated with respect to its analytic value (See
e.g.~\cite{Lousto:2007db}). This means that, although
there may be significant errors in both the magnitude and direction of
the recoil in a particular run, the maximum recoil (as determined by
the fit~(\ref{eq:v_v_th})) for a given $q$ is relatively insensitive
to this error. This can be seen in the dependence of $A$ in
Eq.~(\ref{eq:v_v_th}) on resolution for the $q=1/4$ case in
Table~\ref{table:vofq}. This relatively small difference is
smaller than the mean error in the fits to the form~(\ref{eq:ffit}).
If we perform the fit to form~(\ref{eq:ffit}) using the
high-resolution $q=1/4$ results, then the fitting parameters change
from $(G=62965\pm746, H=2.03\pm0.04)$ to $(G=62908\pm805,
H=2.01\pm0.05)$
and from
$(G=62023\pm555, H=2.13\pm.04)$ to $(G=61974\pm584, H=2.11\pm0.04)$
for the choices $(r_{+}, r_{0}, r_{-})=(2,1,0.5)$ and $(r_{+},
r_{0}, r_{-})=(2.2,1.2,0.7)$
respectively (i.e.\ the change is not statistically significant).

Note that although we expect the $q=1/6$ and $q=1/8$ results to be
less accurate than those of the more modest  mass ratio cases,
including these simulations does not change the observed $\eta$
dependence. This strongly suggests that the $\eta$ dependence is robust
against finite difference errors.

We note that very recent work in PN prediction of the
recoil~\cite{Racine:2008kj} and perturbative calculations (a work in
progress by the authors) show that there is an $\eta^3$ correction to
the leading $\eta^2$ dependence of the out-of-plane recoil. An attempt
to fit our data to the functional form $V/\alpha = (A \eta^2 + B
\eta^3) /(1+q)$ does not yield an accurate evaluation of $B$ because
the functional form of the recoil is insensitive to $B$ at the level
of accuracy we obtained. That is to say, even if $B$ is large, for
example $B\sim 39000$ and $A\sim 52000$ is correspondingly smaller
(such that $B \eta^3 + A \eta^2 \sim 60000/16$ when $\eta=1/4$) the
maximum change in the predicted recoil is of order $10\ \KMS$ over the
entire range of $\eta$. We note that both the PN and perturbative
predictions include a leading order $\eta^2$ dependence
for all configurations including those of Baker et al.~\cite{Baker:2008md}.

\subsection {The In-Plane Recoil}
The empirical formula~(\ref{eq:empirical}) provides an accurate
prediction for the large out-of-plane recoil as a function of mass
ratio (and most importantly, an accurate prediction of the magnitude
of the recoil). Interestingly, the in-plane recoil seems to be larger
than initially expected. Examining Table~\ref{table:transformedkick},
we see that the magnitude of the in-plane recoil $v_{\perp}$ is larger
than the predictions of Eq.~(\ref{eq:empirical}) (the maximum
predicted in-plane recoil velocity for the unequal mass  cases is
about $100$ - $200\ \KMS$ and is dominated by the unequal-mass, rather
than spin, component of the recoil; while for the equal-mass case, the
in-plane recoil is $\sim 50\ \KMS$). The excess in-plane recoil can be
understood  in terms of a newly discovered higher-order effect (here
demonstrated for the first time) that is due to the precession of the
orbital plane.  To understand this effect we need to consider how the
net recoil is generated. As the binary merges, asymmetrical radiation
leads to an oscillation in the momentum of the center of mass. In the
more symmetrical `superkick' configurations, with equal-mass and
equal-and-opposite spins (with spins in the orbital plane), this leads
to the center-of-mass alternately recoiling up and then down in an
oscillatory manner. The total recoil is then determined by where in
this cycle the plunge occurs.  Here this upward/downward oscillation
is superimposed upon a precession of the orbital plane itself. Thus,
for example, a strong `upward' recoil (i.e.\ a recoil along the
orbital angular momentum direction) at one point during the merger
will not be exactly in the opposite direction as  the following
`downward' recoil. This has two effects, the two recoils can never
cancel exactly, and a recoil that initially was entirely out-of-plane
actually has an in-plane component with respect to the final plunge
orbital plane.  Thus both the `upward' and following `downward' recoil
both contribute, in general, to the in-plane recoil.  To put it
another way, in the precessing case, there is a correction to the
in-plane-recoil which is a function of the history of the
instantaneous out-of-plane recoil and the extent that the orbital
plane precesses when the instantaneous recoil is large.  However, the
recoil is only large near merger, and the orbital plane does not
precess to a high degree during this short time period (making the
correction a higher-order effect). Thus the effect, while significant
for the much smaller in-plane recoil, is not significant for the much
larger out-of-plane recoil.  In addition, we note that, as seen in
Fig.~\ref{fig:r2105_fit} the empirical formula underestimates the
out-of-plane recoil to a moderate degree, which is consistent with the
non-cancellation of the upward/downward recoils induced by the orbital
plane precession. (We note that the amount of precession, per orbit,
becomes progressively smaller for larger orbital separations, and hence
the cumulative effect of this non-cancellation is not expected to
introduce a significant recoil at the orbital separations
corresponding to the start of our simulations.) Thus it appears that
there are higher-order effects in the precessing case that perturb the
out-of-plane recoil as well.  These newly discovered effects are
likely modeled by higher-order PN terms in the empirical formula and
we are in the process of investigating how the in- and out-of-plane
recoils are affected, with the goal of incorporating these effects
into the formula.  This will be the subject of a forthcoming paper by
the authors.

We also note that the that the difference in the magnitude of the
in-plane component of the recoil for the two choices of $(r_{+},
r_{0}, r_{-})$ are around $25\%$. These differences are
due to the fact that the out-of-plane recoil is very large compared to
the in-plane recoil, and hence a small error in the direction
associated with the orbital plane will lead to relatively large
changes in the calculated `in-plane' recoil. For example, if the
out-of-plane recoil is $1000\ \KMS$ and the direction of the orbital
plane has an error of $5^\circ$, then the in-plane recoil will have
an error of $1000\ \sin(5^\circ)\ \KMS \sim 90\ \KMS$. Note that this
error is too small to explain the large magnitude of the in-plane
recoil, which appears to be due to a newly discovered physical effect
mentioned above.

\subsection {An alternative choice for the plane}
\label{sec:alt_plane}
\begin{figure}
\caption{$M d^2r/dt^2$ versus $r/M$ as a function of mass ratio and initial
spin orientation. Note that the locations of the zero, minima, and maxima
are very similar.}
\includegraphics[width=3.0in]{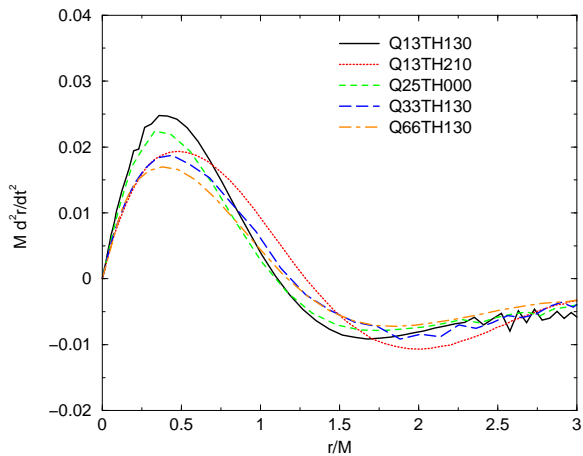}
\label{fig:ddotr}
\end{figure}

\begin{figure}
\caption{Fit of out-of-plane recoil ($V_\|/\alpha$) versus mass ratio
$q$ for the adaptive choice of $(r_{+}, r_{0}, r_{-})$.  The top
shaded region shows our predicted value for $K=(6.0\pm0.2)\times
10^4$, the lower region shows the prediction based on the modification
proposed by Baker et al., the lower (green) curve shows a fit to
$\eta^n$, while the upper (blue) curve has a simultaneous fit to
$\eta^n$ and $K$. Note that the measured values of $V_\|/\alpha$,
while slightly overshooting the predictions, agree much better with
our ${\cal O}(\eta^2)$ form. The dots are the locations of the data
points. }
\includegraphics[width=3.0in]{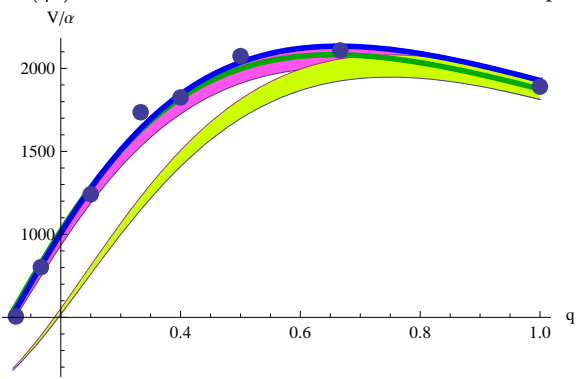}
\label{fig:ref_fit}
\end{figure}

In this section we reanalyze the recoil velocities using a choice of
orbital plane adapted to each individual run. In Fig.~\ref{fig:ddotr}
we show $\ddot r(t)$ as a function of $r$ during the merger. From the
figure we can see that there is some scatter in the locations of the
zero, maximum, and minimum of $\ddot r(t)$, but are generally
independent of both the initial angle and mass ratio (this observation
justifies our choice above of using fixed choices for $(r_+, r_0,
r_-)$). In this analysis we chose $r_+$ to be the location of the
minimum in $\ddot r(t)$, $r_0$ to be the location of the zero (i.e.\
location where $|\dot r(t)|$ is maximized), and $r_-$ to be the
location of the maximum in $\ddot r(t)$. This choice has the advantage
that we do not use any `fiducial' choices for $(r_+, r_0, r_-)$.
Table~\ref{table:ref_transformedkick} summarizes the results for the
transformed recoil velocities and kicks, while
Table~\ref{table:ref_vofq} summarizes the fit parameters $A$ and $B$
after fitting the results from each mass ratio to the
form~(\ref{eq:v_v_th}).  A fit of the maximum recoil velocity per mass
ratio to the form~(\ref{eq:ffit}) yields $H=1.904\pm0.041$ when $G=K$
and ($H=1.954\pm0.05$, $G=(6151\pm804) \KMS$) when both $H$ and $G$
are allowed to vary. These results are in good agreement with those
for the fixed choices of $(r_+, r_0, r_-)$. Finally, in
Fig.~\ref{fig:ref_fit} we show fits of the maximum out-of-plane recoil
versus mass ratio using this choice of adaptive plane parameters.
Note that with this method the scatter in the plot is more significant,
indicating that a fixed choice of plane parameters produces a more robust
estimate of the out-of-plane recoil. We also note that the choice of plane
parameters $(r_+,r_0,r_0) = (2.2, 1.2, 0.7)$ gives the best fits, which is consistent
with the observation that $r_0=1.2$ is well adapted to the locations
of the maximum in $|\dot r|$  for
the runs considered here.
\begin{table}
\caption{The transformed spin (at $r=r_0$), recoil velocities in
$\KMS$, and angles between the in-plane QXXXTH000 spins and the other
QXXXTHYYY configurations in degrees when $r_+$ is chosen to be the
value of $r$ when $\ddot r(t)$ reaches its minimum, $r_0$ is the value
of $r$ when $\ddot r(t)=0$ (i.e. when $|\dot r(t)|$ is at its
maximum), and $r_-$ is the value of $r$ when $\ddot r(t)$ reaches its
maximum.  Note that in the transformed system $L_z$ and
$S_z$ are both negative (i.e.\ there is some partial spin/orbit
alignment. We denote quantities in the (transformed) orbital plane
with a $\perp$ subscript.}
  \label{table:ref_transformedkick}
\begin{ruledtabular}
\begin{tabular}{lccccc}
Config   & $V_{\perp} $ & $V_{\|} $ & $a_{\perp}/m^H $ & $a_{\|}/m^H $ & $\vartheta $ \\ \hline
Q13TH000 & 52.1681 & -112.324 & 0.770573 & -0.236828 & 0. \\
Q13TH090 & 58.7557 & -101.697 & 0.7674 & -0.246703 & 1.29153 \\
Q13TH130 & 213.996 & -372.167 & 0.715948 & -0.357787 & -55.7042 \\
Q13TH210 & 37.8511 & 130.01 & 0.764949 & -0.244748 & -178.98 \\
Q13TH315 & 234.147 & 372.257 & 0.733002 & -0.319972 & 105.641 \\
\\
Q17TH000 & 278.533 & 584.276 & 0.73713 & -0.307434 & 0. \\
Q17TH090 & 133.506 & 345.022 & 0.747727 & -0.292225 & 56.9267 \\
Q17TH130 & 405.804 & 305.157 & 0.796759 & -0.0790985 & -66.0061 \\
Q17TH210 & 273.036 & -577.037 & 0.729478 & -0.32118 & -170.825 \\
Q17TH315 & 303.864 & -110.066 & 0.797954 & -0.0710164 & 87.869 \\
\\
Q25TH000 & 328.136 & -965.098 & 0.74441 & -0.277851 & 0. \\
Q25TH090 & 368.549 & -909.203 & 0.767864 & -0.203131 & -20.4251 \\
Q25TH130 & 194.765 & 59.5598 & 0.777913 & -0.186817 & -111.668 \\
Q25TH210 & 340.595 & 951.677 & 0.758298 & -0.24157 & 167.113 \\
Q25TH315 & 122.309 & -205.964 & 0.771452 & -0.210448 & 73.1503 \\
\\
Q33TH000 & 432.797 & -1210.91 & 0.784273 & -0.156319 & 0. \\
Q33TH090 & 522.116 & -865.374 & 0.797412 & -0.0611074 & -36.3755 \\
Q33TH130 & 123.321 & 392.695 & 0.764043 & -0.242325 & -87.9496 \\
Q33TH210 & 538.156 & 1015.2 & 0.798003 & -0.0555484 & 148.237 \\
Q33TH315 & 160.341 & -603. & 0.755614 & -0.263486 & 78.0326 \\
\\
Q40TH000 & 345.769 & -1460.64 & 0.777363 & -0.199695 & 0. \\
Q40TH090 & 464.445 & -1240.18 & 0.795298 & -0.0996155 & -31.0583 \\
Q40TH130 & 199.788 & 84.8549 & 0.779402 & -0.184948 & -97.0931 \\
Q40TH210 & 358.115 & 1413.71 & 0.783135 & -0.174788 & 177.471 \\
Q40TH315 & 138.301 & -323.804 & 0.77561 & -0.207095 & 73.4494 \\
\\
Q50TH000 & 206.224 & 30.7051 & 0.780222 & -0.185459 & 0. \\
Q50TH090 & 216.108 & 1263.93 & 0.77053 & -0.230075 & -47.8033 \\
Q50TH130 & 351.439 & 1578.89 & 0.789517 & -0.13106 & -98.1525 \\
Q50TH210 & 122.003 & -618.748 & 0.768792 & -0.235123 & 158.316 \\
Q50TH315 & 379.77 & -1493.21 & 0.797621 & -0.0947157 & 68.7318 \\
\\
Q66TH000 & 134.205 & 890.868 & 0.769565 & -0.233273 & 0. \\
Q66TH090 & 253.359 & 1717.31 & 0.797964 & -0.106154 & -50.8354 \\
Q66TH130 & 324.629 & 1042.76 & 0.800428 & -0.0583467 & -100.477 \\
Q66TH210 & 164.115 & -1364.25 & 0.769748 & -0.233049 & 171.955 \\
Q66TH315 & 296.905 & -840.357 & 0.799556 & -0.0682437 & 72.8232 \\
\\
Q100TH000 & 89.5687 & 1423.6 & 0.793956 & -0.16161 & 0. \\
Q100TH090 & 119.496 & 1015.99 & 0.804421 & -0.0421833 & -66.5996 \\
Q100TH130 & 52.4861 & -244.07 & 0.789395 & -0.14723 & -119.948 \\
Q100TH210 & 80.6623 & -1485.43 & 0.803926 & -0.0816034 & 151.372 \\
Q100TH315 & 60.4502 & 387.967 & 0.788101 & -0.157316 & 56.6673 \\
\end{tabular}
\end{ruledtabular}
\end{table}
\begin{table}
\caption{Fit parameters for $v_\| = A \cos((\vartheta - B)\pi/180)$ for each $q$.
$A$ is in units of $\KMS$ and $B$ is in degrees for the adapted choice
of plane parameters $(r_+, r_0, r_-)$.}
\label{table:ref_vofq}
\begin{ruledtabular}
\begin{tabular}{lcc}
$q$ & A & B \\
\hline
1/8 & $379.0\pm9.9$ & $108.51\pm1.33$\\
1/6 & $611.4\pm37.4$ & $353.47\pm3.33$\\
1/4 & $948.5\pm55.6$ & $169.16\pm4.08$\\
1/3 & $1353.4\pm66.5$ & $195.32\pm2.34$\\
1/2.5 & $1427.5\pm15.6$ & $176.70\pm0.72$\\
1/2 & $1620.5\pm16.7$ & $271.46\pm0.60$\\
2/3 & $1659.9\pm82.8$ & $310.17\pm3.05$\\
1 & $1504.9\pm15.3$ & $340.60\pm0.56$\\
\end{tabular}
\end{ruledtabular}
\end{table}

\begin{widetext}

\begin{figure}
\caption{The recoil velocity versus time for various mass ratios for
the $\Theta=0$ configurations (top left: $q=1/8$, top right: $q=1/4$,
bottom left: $q=2/3$, bottom right: $q=1$). The error in the recoil
due to not including the correct initial value has the effect of
translating the plots. This error appears to be no more than $\sim10\ \KMS$.
}
\includegraphics[width=3.0in]{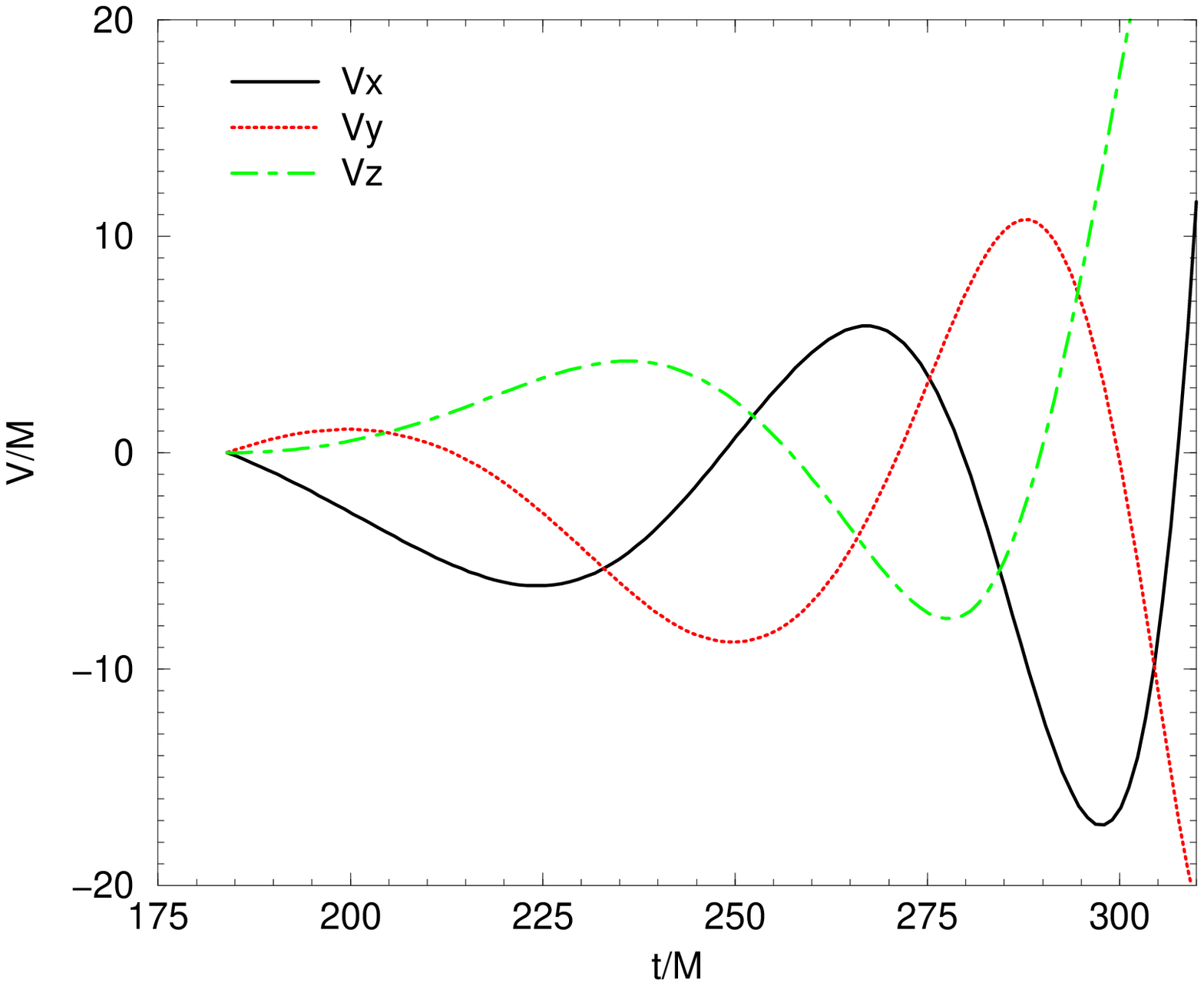}
\includegraphics[width=3.0in]{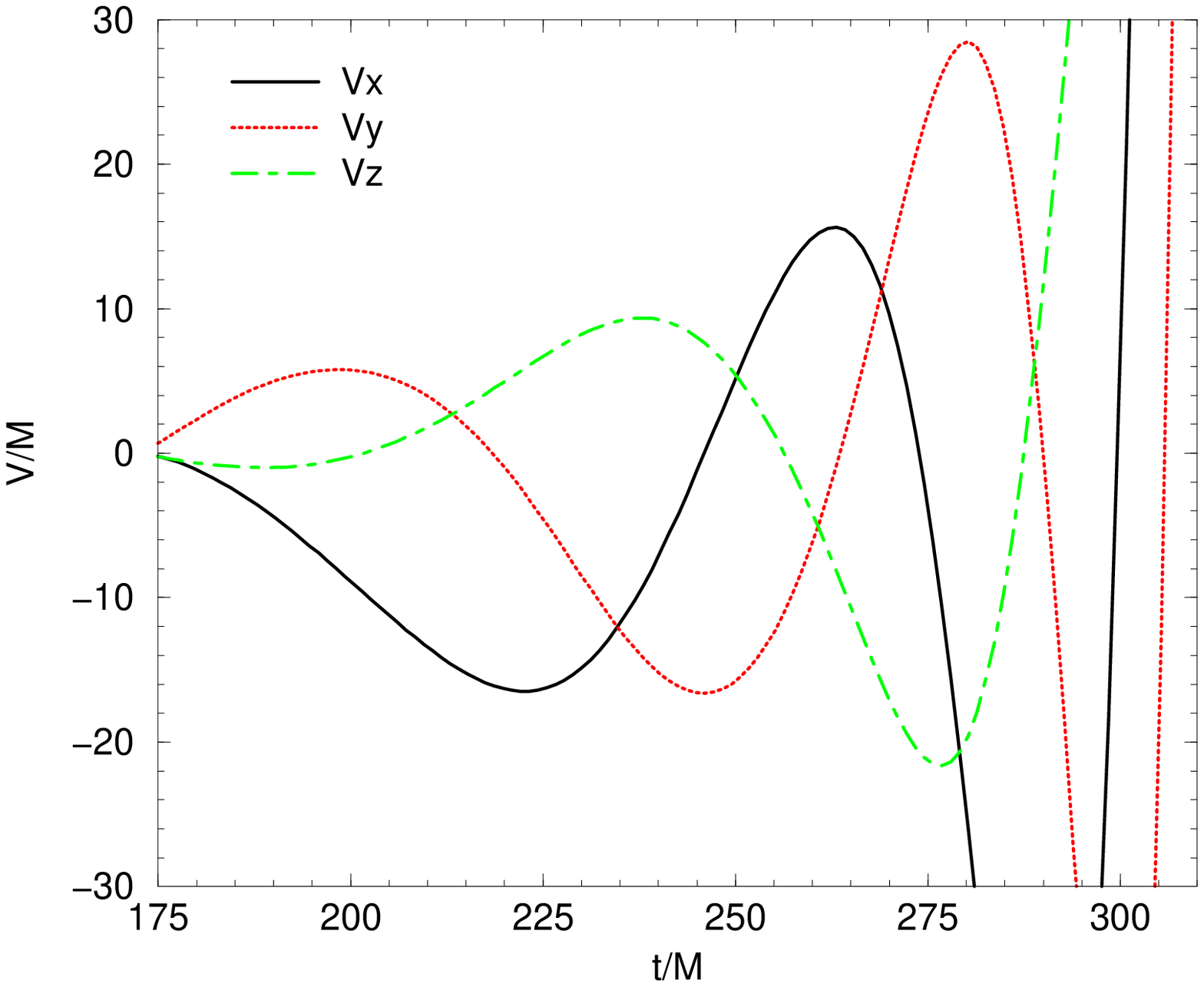}
\includegraphics[width=3.0in]{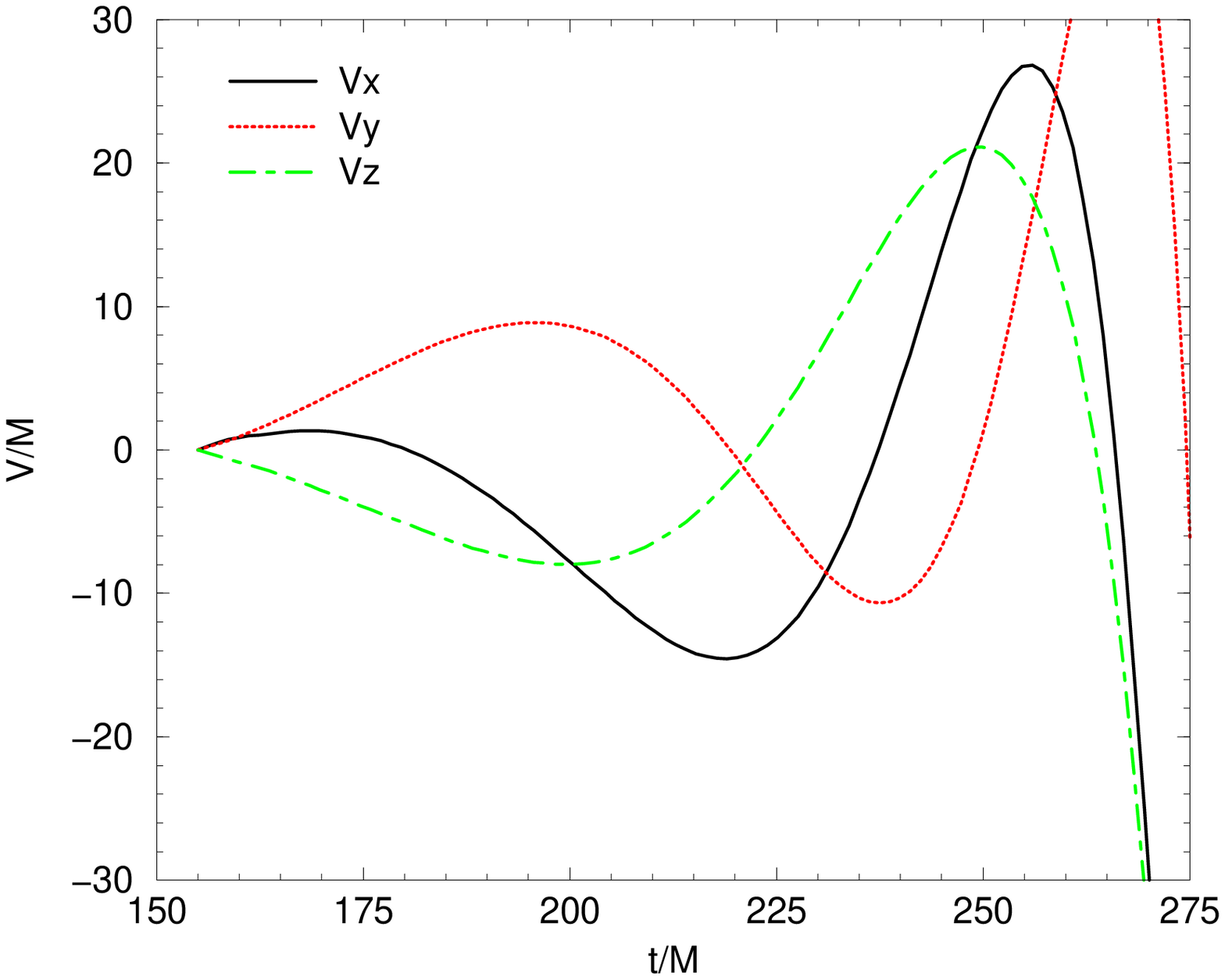}
\includegraphics[width=3.0in]{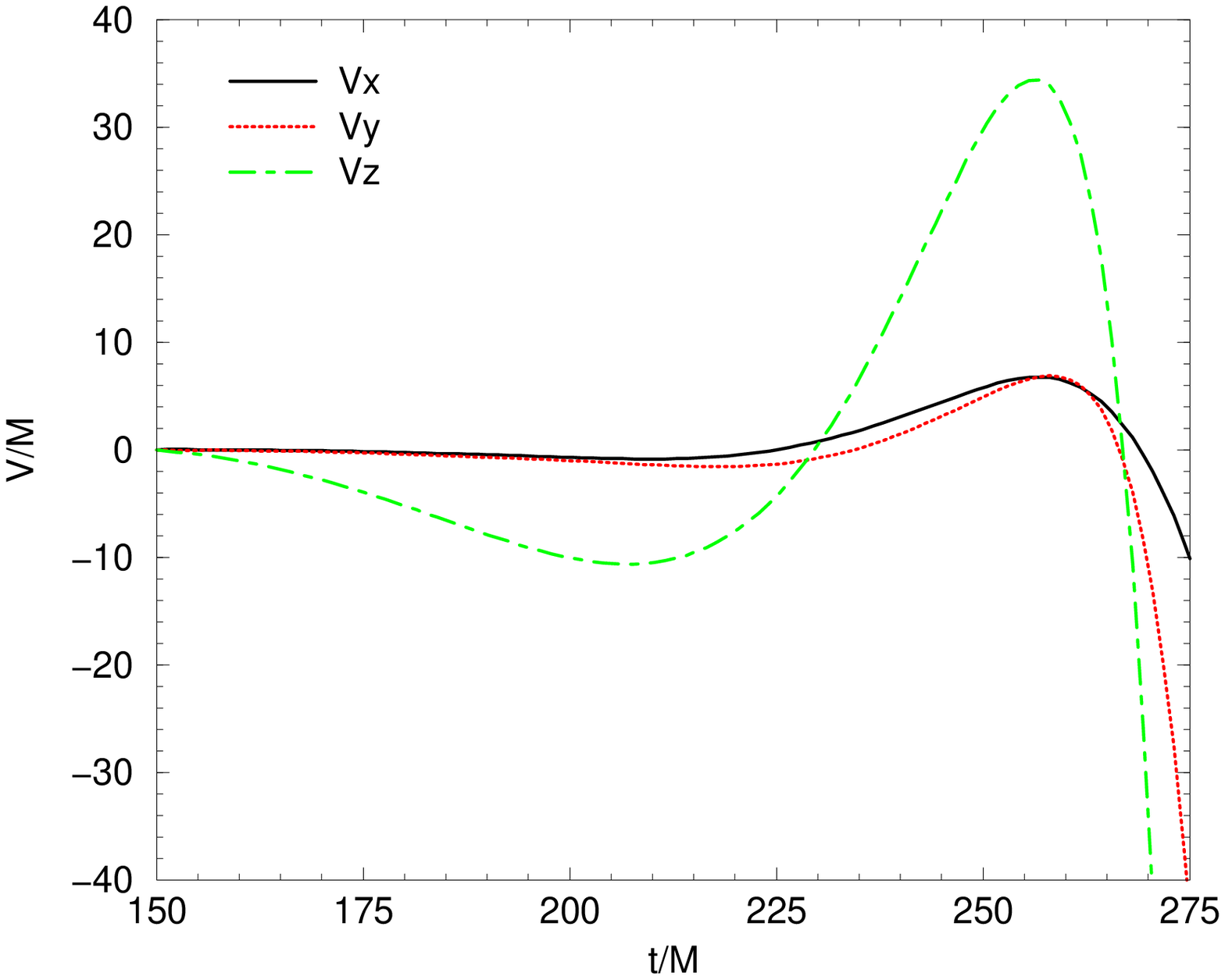}
\label{fig:in_kick}
\end{figure}
\end{widetext}

Finally, we note that the error in the recoil velocities due to
finite starting time were of order
$10\ \KMS$ in each direction for all runs. We estimated these
errors by looking for a systematic offset in the recoil velocity
when plotted as a function of time (See Fig~\ref{fig:in_kick}).
These errors, while not
insignificant do not change the results of our analysis because they
are the same order as the errors due to extrapolations to infinity and
the errors in the fits of the recoil. 

\section{Conclusion}
\label{sec:discussion}
In this paper we explored the merger recoil from precessing black-hole
binaries with a larger spinning black hole (with initial spin $a/m =
0.8$) in quasi-circular orbit with a smaller non-spinning black hole.
We introduced techniques to determine the normal to the orbital plane
at merger and thus decompose the recoil into its in-plane and out-of
plane components. However, there are important open questions about
the accuracy of the determination of the orbital plane, the spin
direction, and spin magnitude. All these quantities are measured in
the highly dynamical region around the merger. In this work we have
introduced techniques to begin studying this problem that will need to
be refined. The issue of the in-plane recoil is particularly
important, because large in-plane recoils imply that our heuristic
formula needs to be modified for strongly precessing binaries.
Nevertheless, our results, as seen in Fig.~\ref{fig:fit}, indicate
that the out-of-plane recoil has  an ${\cal O} (\eta^2)$ rather than
${\cal O}(\eta^3)$ leading-order dependence on the symmetric mass
ratio. This result, while agreeing with our prediction, appears to
disagree with the recent work of Baker et al.~\cite{Baker:2008md}.
Thus additional work with new configurations may be needed in order to
determine leading-order dependence, or indeed, if this dependence is a
function of the configuration.

\acknowledgments
We thank Manuela Campanelli for many valuable discussions.
We gratefully
acknowledge NSF for financial support from grant PHY-0722315,
PHY-0653303, PHY 0714388, and PHY 0722703; and NASA for
financial support from grant NASA 07-ATFP07-0158 and
HST-AR-11763.01.
Computational resources were provided by Lonestar cluster at TACC
and by NewHorizons at RIT.

\bibliographystyle{apsrev}
\bibliography{../../../Bibtex/references}

\end{document}